\documentclass[aps,prb,onecolumn,groupedaddress]{revtex4-1}

\usepackage{amsmath}
\usepackage{graphicx}
\usepackage{verbatim}
\usepackage{bm}

\newcommand{\su}{spin-$\uparrow$}

\newcommand{\sd}{spin-$\downarrow$}

\newcommand{\vect}[1]{\mathbf{#1}}

\begin{document}

\title{Effects of Defects and Dephasing on Charge and Spin Currents in Two-Dimensional Topological Insulators}

\author{John S. Van Dyke}
\author{Dirk K. Morr}

\affiliation{University of Illinois at Chicago, Chicago, Illinois 60607, USA}

\date{\today}

\begin{abstract}
Using the non-equilibrium Keldysh Green's function formalism, we investigate the effect of defects on the electronic structure and transport properties of two-dimensional topological insulators (TI). We demonstrate how the spatial flow of charge changes between the topologically protected edge and bulk states and show that elastically and inelastically scattering defects that preserve the time reversal symmetry of the TI lead to qualitatively different effects on the TI's local electronic structure and its transport properties. Moreover, we show that the recently predicted ability to create highly spin-polarized currents by breaking the time-reversal symmetry of the TI via magnetic defects [Phys. Rev. B {\bf 93}, 081401 (2016)] is robust against the inclusion of a Rashba spin-orbit interaction and the effects of dephasing, and remains unaffected by changes over a wide range of the TI's parameters. We discuss how the sign of the induced spin currents changes under symmetry operations, such as reversal of bias and gate voltages, or spatial reflections. Finally, we show that the insight into the interplay between topology and symmetry of the magnetic defects can be employed for the creation of novel quantum phenomena, such as highly localized magnetic fields inside the TI.

\end{abstract}

\pacs{}

\maketitle

\section{Introduction}
\begin{figure}[h!]
 \begin{center}
\includegraphics[width=9cm]{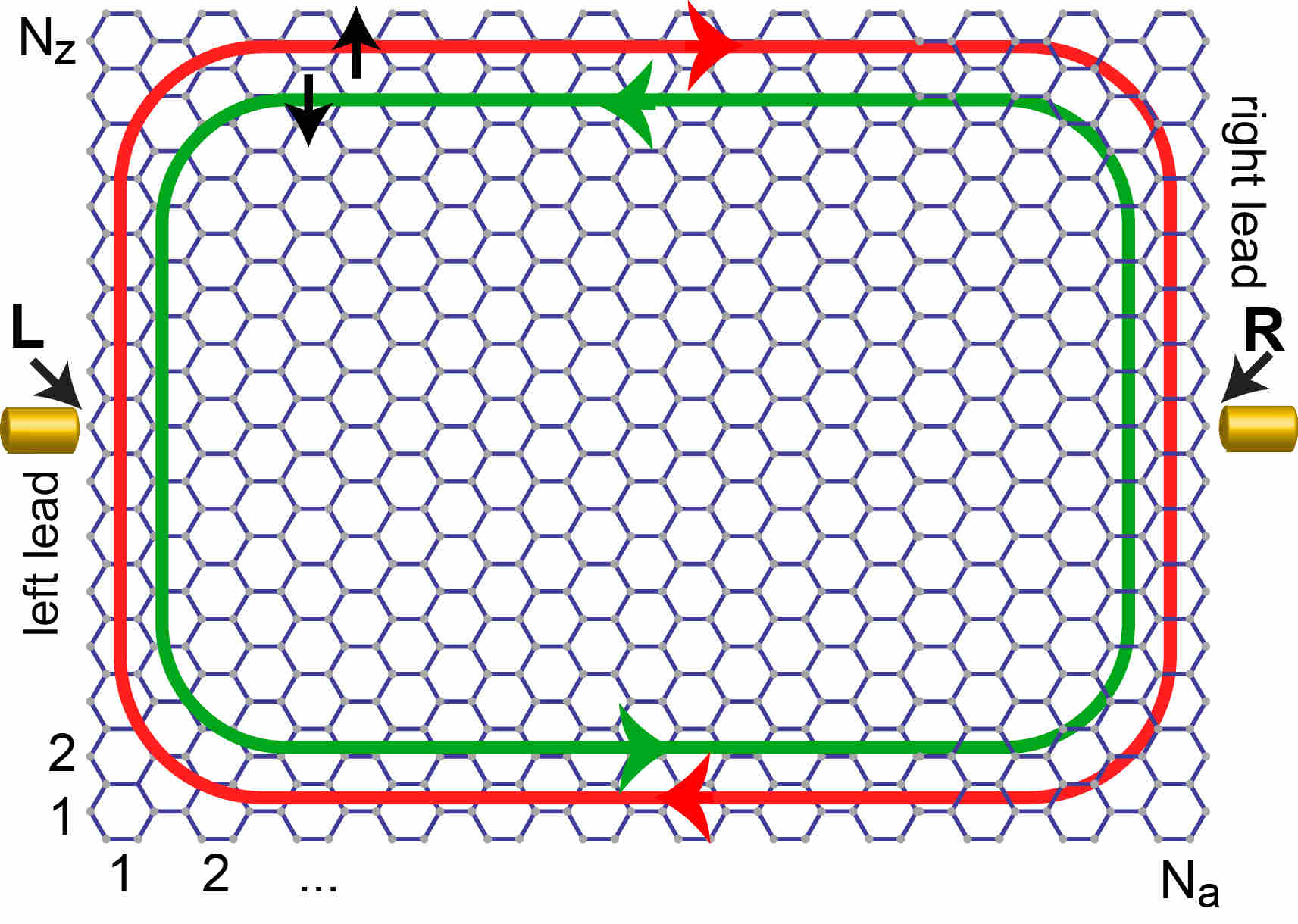}
\caption{Schematic representation of the spin-resolved spatial current patterns in a two-dimensional TI with weak coupling to the leads.} \label{fig:TI_lattice}
\end{center}
\end{figure}

Topological insulators (TIs) have attracted great interest over the last decade \cite{Hasan2010,Moore2010,Qi2011,Ando2013}, not only because they represent an intriguing state of matter whose properties are determined by topology, but also because of their possible applications \cite{Yok09,Akh09} in fields ranging from spintronics \cite{Wolf2001} to quantum computation \cite{Kane1998}.  TIs are characterized by non-zero topological invariants \cite{Moore2007} that reflect the coexistence of gapless edge or surface states with insulating bulk states. As such, considerable efforts have focused on the classification of the topological states and their possible realization in experiment \cite{Altland1997,Ryu2010,Hasan2010,Qi2011}. A crucial element of TIs is the presence of a spin-orbit interaction, which in two-dimensional (2D) topological insulators \cite{Rasche2013,Konig2007,Roth2009,Kane2005b,Young2014} leads to the existence of helical edge states \cite{Kane2005a,Kane2005b}. These edge states represent Kramers doublets of counterpropagating states with opposite spin polarization, as shown in Fig.~\ref{fig:TI_lattice}.
This unique locking of momentum and spin immediately implies that electrons can only scattered between these helical edge states through a process that involves backscattering in combination with a spin-flip. It is this requirement that renders these helical edge states robust against any defects that preserve the time-reversal symmetry of the system \cite{Kane2005a}, such as elastically scattering potential defects, as well as against ensuing localization effects \cite{Anderson1958}. The resulting robustness of the topologically protected edge states has rendered topological insulators of great interest for any applications \cite{Yok09,Akh09} that require long coherence times, as are often found in spintronics \cite{Wolf2001}.

On the other hand, it has been shown that if the time reversal symmetry in a TI is broken \cite{Chang2014,Yu2010,Qi2008b} for instance by using magnetic defects \cite{Liu2009,Chen2010}, intriguing new phenomena can occur, such as a topological magnetoelectric effect \cite{Qi2008a,Essin2009}, a quantum anomalous Hall effect \cite{Qi2008a,Liu2008,Yu2010}, and image magnetic monopoles \cite{Qi2009}. We recently demonstrated \cite{VanDyke2016} that the breaking of the time reversal symmetry in 2D TIs via magnetic defects or in magnetic heterostructures can be employed for the creation of nearly perfectly spin-polarized currents, as well as highly tunable spin diodes. Experimental evidence for the existence of spin polarization of current in three-dimensional TIs up to room temperature was recently reported in Refs. \cite{Li2014,Tian2014,Tang2014,Dankert2015,Liu2015}. It is therefore the combination of long coherence times and the ability to create spin-polarized currents that might hold the key for employing TIs in the next generation of spintronics devices.

As the realization of these applications will likely occur in TIs on the sub-100nm scale \cite{Xiu2011}, several important questions arise. First, to what extent do defects that preserve the time reversal symmetry of the TI, such as elastically scattering non-magnetic impurities or molecules with vibrational (phonon) modes, change the electronic structure of the TI and its transport properties?  Second, how robust is the predicted ability to create a large spin polarization of currents in TIs \cite{VanDyke2016} against variations in parameters of the TI often encountered in real systems, such as a varying scattering strength of magnetic defects, the particular locations of defects, aspect ratio and the size of the TI, or width of the leads?  In particular, to what extent can the spin polarization be destroyed by the inclusion of (a) dephasing induced by the interaction with phonons, or (b) a Rashba spin-orbit interaction? Third, how does the spin polarization of currents change under symmetry operations, such as reversal of the bias and gate voltages, or spatial reflections around symmetry points or lines? This question is of particular interest for the envisioned creation of spin-diodes. Fourth, can one employ the interplay between symmetry of the magnetic defects and the topological structure of the TI for the creation of novel quantum phenomena?

We will address all of these questions in this article. In particular, we will show that the interaction with phonon modes (albeit preserving the TI's time-reversal symmetry) can not only qualitatively change the TI's transport properties, but can lead to a destruction of the topological nature of its helical edge states. Moreover, we will demonstrate that the predicted creation of highly spin-polarized currents is robust and does not, for example, depend on a particular size of the TI, or specific parameters for the magnetic scattering strength. In particular, the phonon-induced dephasing effects or a Rashba interaction can only destroy the ability to create spin-polarized currents to the extent that they also destroy the topological nature of the TI itself. Finally, we will provide an example of how the newly gained insight into the interplay between topology and symmetry of the magnetic scattering can be employed for the creation of a novel quantum phenomenon, the formation of highly localized and tunable magnetic fields around interior holes in the TI.

This paper is organized as follows. In Sec.~\ref{sec:theory} we provide a brief review of the non-equilibrium Keldysh formalism employed here and in Ref.~\cite{VanDyke2016}  to compute charge and spin currents in two-dimensional TIs. In Sec.~\ref{sec:evol} we discuss how the local density of states and the spatial current patterns in a clean 2D TI evolve as one transitions from the helical edge states to the bulk states. In Sec.~\ref{sec:potdefects} we discuss the effects of  time reversal symmetry-preserving defects, including non-magnetic impurities and local phonon modes, on the electronic structure and transport properties of the TI. In Sec.~\ref{sec:magdefects} we show that the creation of high spin-polarized currents using magnetic defects is robust against variations in the magnetic scattering strength, the number and positions of impurities, etc. In Sec.~\ref{sec:symmetries} we discuss the symmetries of
the symmetry properties of the spin polarization under transformations of the bias or gate voltages, spatial transformations, and changes in the signs of the scattering strength. In Sec.~\ref{sec:magfield} we demonstrate that the newly gained insight into the effects of magnetic impurities can be employed to create novel quantum phenomena, such as highly localized magnetic fields. Finally, in Sec.~\ref{sec:concl} we summarize our results and present our conclusions.

\section{Theoretical Formalism}
\label{sec:theory}

In this section, we briefly outline the non-equilibrium Green's function formalism \cite{Keldysh1965,Caroli1971} that we employ to compute spin-resolved currents in a two-dimensional topological insulator. We also present some analytical results for the changes in the energy of edge states due to the scattering off defects.

\subsection{Charge and Spin Transport}
\label{sec:theoryI}

Our starting point for the study of finite two-dimensional topological insulators with a hexagonal (graphene-like) lattice structure is the Kane-Mele Hamiltonian \cite{Kane2005a}
\begin{align}
H=&-t\sum_{\langle {\bf r,r'}\rangle,\alpha} c^\dagger_{ {\bf r},\alpha} c_{ {\bf r'},\alpha} + i \Lambda_{SO} \sum_{ \langle \langle {\bf r,r'}\rangle \rangle,\alpha, \beta}
 \nu_{\bf r,r'} c^\dagger_{ {\bf r},\alpha} \sigma_{\alpha \beta}^z  c_{ {\bf r'},\beta} \notag \\
 &+ i \Lambda_R \sum_{\langle {\bf r,r'} \rangle} c^\dagger_{ {\bf r},\alpha} ( \boldsymbol{\sigma} \times \hat{{\bf d}}_{{\bf r r'}} )^z_{\alpha \beta} c_{ {\bf r'},\beta}
 - t_{l} \sum_{ {\bf r,r'} ,\alpha} \left( d^\dagger_{ {\bf r},\alpha} c_{ {\bf r'},\alpha} + h.c. \right) + H_{lead}
\label{eq:fullH}
\end{align}
where the first three terms on the right-hand-side represent the conventional electronic hopping between nearest-neighbor sites, the spin-orbit induced hopping between next-nearest neighbor sites (with $\nu_{\bf r,r'}=-\nu_{\bf r,r'}=\pm 1$, and $\sigma_{\alpha \beta}^z$ being a Pauli matrix), and the Rashba spin-orbit interaction that results in an electron's spin-flip when hopping  between nearest-neighbor sites (with $\boldsymbol{\sigma}$ a vector of Pauli matrices and $\hat{{\bf d}}_{{\bf r r'}}$ a unit vector in the direction connecting sites ${\bf r}$ and ${\bf r'}$).  The fourth term represents the hopping between the TI and the leads, respectively. Here, $c^\dagger_{ {\bf r},\alpha}, c_{ {\bf r},\alpha}$ are the fermionic creation and annihilation operators, creating or annihilating an electron with spin $\alpha$ at site ${\bf r}$. Similarly, $d^\dagger_{ {\bf r},\alpha}, d_{ {\bf r},\alpha}$ creates or annihilates an electron  with spin $\alpha$ at site ${\bf r}$ in the leads. $H_{lead}$ describes the electronic structure of the leads, which, however, is largely irrelevant for the TI's transport properties.

The scattering of the TI's conduction electrons by non-magnetic defects is described by the Hamiltonian
\begin{equation}
H_{pot} = U_0 {\sum_{{\bf R},\alpha}} c^\dagger_{{\bf R},\alpha} c_{{\bf R},\alpha}
\label{eq:U0}
\end{equation}
where $U_0$ is the non-magnetic (potential) scattering strength, and the sum runs over all defect locations (we assume point-like scatterers). Similarly, the scattering by magnetic defects is described by the Hamiltonian \cite{Liu2009,Chen2010}
\begin{equation}
H_M={\sum_{\bf R}} J_z S^z_{\bf R} \left( c^\dagger_{ {\bf R},\uparrow} c_{ {\bf R},\uparrow} - c^\dagger_{ {\bf R},\downarrow} c_{ {\bf R},\downarrow} \right) + J_\pm \left( S^+_{\bf R} c^\dagger_{ {\bf R},\downarrow} c_{ {\bf R},\uparrow} + S^-_{\bf R} c^\dagger_{ {\bf R},\uparrow} c_{ {\bf R},\downarrow} \right)
\label{eq:Jscatt}
\end{equation}
where the sum runs over all defect locations. We assume the magnetic defects to be static in nature, implying that $S^{z,\pm}_{\bf R}$ simply becomes a $c$-number with $J_z S^z_{\bf R} = J_z S$, $J_\pm S^+_{\bf R} = J_\pm S (1+i)$, and $J_\pm S^-_{\bf R} = J_\pm S (1-i)$. This assumption can be justified by the fact that the Kondo temperature $T_K$ \cite{Wu2006,Maciejko2009} can be strongly suppressed either by the absence of edge states near the Fermi energy \cite{Rossi2006}, the use of large-spin defects, or by applying local static magnetic fields \cite{Rugar2004}. On the other hand, the topological nature of TIs can persist up to room temperature \cite{Dankert2015}, such that there exists a sufficiently large temperature range above $T_K$ in which the magnetic defects can be considered static \cite{Chen2010}.

To investigate the flow of charge and spin in a finite, two-dimensional TI, we employ the non-equilibrium Keldysh Green's function formalism \cite{Keldysh1965,Caroli1971}. Within this formalism,
the spin-resolved current between sites ${\bf r}$ and ${\bf
r}^\prime$ in the TI is induced by different
chemical potentials, $\mu_{L,R}=\pm V_0/2$ in the left and right
leads, and given by \cite{Caroli1971}
\begin{equation}
I^\sigma_{\bf r  r^\prime}=-2 \frac{e}{\hbar} \;
\intop_{-\infty}^{+\infty}\frac{d\omega}{2\pi}{\rm Re} \left[ t_{\bf r  r^\prime}
G^<_{\bf r  r^\prime}(\sigma, \omega)\right] \ . \label{eq:Current}
\end{equation}
with $\sigma = \uparrow, \downarrow$ representing the spin degrees of freedom, $t_{\bf rr'}^\sigma$ being the real ($-t$) or imaginary ($\pm i\Lambda_{SO}$, $ \pm i\Lambda_{R}$) electron hopping elements between sites ${\bf r}$ and ${\bf r^\prime}$, and $G^<_{\bf r  r^\prime}(\sigma, \omega)$ being the full, spin-resolved non-local lesser Green's function, defined via ${\hat
G}^<_{\bf r  r^\prime}(t,t) = \langle c^\dagger_{\bf r^\prime}(t)  c_{\bf r}(t) \rangle $ in the time domain. The charge current is then given by $I_{out}^c=I_{out}^\uparrow+I_{out}^\downarrow$, and the spin-$\sigma$ polarization of the outgoing current is defined via $\eta_\sigma=I_{out}^\sigma / I_{out}^c$.

To account for the effects of  electronic hopping, the presence of magnetic or non-magnetic defects, the electron-phonon interaction, and the coupling to the leads, we employ the Dyson equations for the lesser and retarded Green's functions. By defining lesser and retarded Green's function matrices $\hat{G}^{<,r}$ in real space whose $({\bf r r'})$  elements are given by $\hat{G}^{<,r}_{\bf r r'}$, we obtain the Dyson equations in frequency space
\begin{subequations}
\begin{align}
\hat{G}^{<} &= \hat{G}^{r}\left[ \left(\hat{g}^{r} \right)^{-1} \hat{g}^{<}
\left( \hat{g}^{a} \right)^{-1} + {\hat \Sigma}^<_{ph} \right] \hat{G}^{a} \label{eq:fullGa} \\
\hat{G}^{r} &= \hat{g}^{r} + \hat{g}^{r} \left[ \hat{t} + {\hat \Sigma}^r_{ph}
\right] \hat{G}^{r}
\label{eq:fullGb}
\end{align}
\end{subequations}
where $\hat{t}$ is the hopping matrix which includes both the real and imaginary hopping elements of Eq.(\ref{eq:fullH}), as well as the scattering strength $U_0$ from Eq.(\ref{eq:U0}) and $J_{z,\pm} S$ from Eq.(\ref{eq:Jscatt}). Moreover, ${\hat \Sigma}^{r,<}_{ph}$ are the retarded and lesser fermionic self-energy matrices arising from the electron-phonon interaction, and $\hat{g}^{r,a,<}$ are the retarded, advanced and lesser fermionic Green's function matrices of the TI and the leads in the absence of any electronic hopping, defect scattering or electron-phonon interaction. These Green's functions are given by $(x=r,a,<)$
\begin{equation}
\hat{g}^{x}= \left(
\begin{array}{cc}
\hat{g}_{leads}^{x} & 0  \\
0 & \hat{g}_{TI}^{x}
\end{array}%
\right)
\end{equation}
where $\hat{g}_{TI}^{x}$ and $\hat{g}_{leads}^{x}$ are the Green's function matrices describing the TI and the right and left leads, respectively. Moreover, $\hat{g}_{TI}^{x}$ are diagonal matrices with elements
\begin{subequations}
\begin{align}
g_0^r(\omega) &= \frac{1}{\omega + i \delta - eV_g} \\
g_0^<(\omega) &= -2i n_F(\omega) {\rm Im} g_0^r(\omega)
\end{align}
\end{subequations}
where $n_F(\omega)$ is the Fermi distribution function, $e$ is the electron charge and $V_g$ is the gate voltage. Note that to move a state from energy $E_i>0$ to the Fermi energy, one has to apply the gate voltage $V_g=E_i/e$. Moreover, $\hat{g}_{leads}^{x}$ are diagonal matrices with elements
\begin{subequations}
\begin{align}
g_{leads}^r(\omega) &= -i \pi \\
g_{leads}^<(\omega) &= -2i \  n_F(\omega+\mu_{L,R}) \ {\rm Im} g_0^r(\omega)
\end{align}
\end{subequations}
implying that the leads' density of states is equal to unity and that we consider the wide band limit for the leads. Moreover, $\mu_{L,R}$ is the chemical potential in the left and right leads, giving rise to a potential difference $\Delta V=(\mu_L - \mu_R)/e$ across the TI. The spin-resolved local density of states, $N_\sigma({\bf r}, E)$ at site {\bf r} and energy $E$ is obtained from Eq.(\ref{eq:fullGb}) via
\begin{equation}
N_\sigma({\bf r}, E=\hbar \omega) = -\frac{1}{\pi} {\rm Im} \ {\hat G}^r_{\bf rr}(\omega) \ .
\end{equation}

To study how the electronic structure of the TI and the spatial current patterns are affected by the interaction of individual sites with molecules containing phonon modes,
or by dephasing arising from the coupling of all TI sites to local phonon modes, we consider the electron-phonon interaction
\begin{equation}
H_{e-ph}= g {\sum_{\textbf{r},\sigma}} c^\dagger_{\textbf{r},\sigma} c_{\textbf{r},\sigma}  \left( a^\dagger_\textbf{r} + a_\textbf{r} \right) + {\sum_\textbf{r}} \omega_0 a^\dagger_\textbf{r} a_\textbf{r} \ , \label{eq:eph}
\end{equation}
where $g$ is the electron-phonon coupling, $a^\dagger_\textbf{r}, a_\textbf{r}$ creates or annihilates a phonon at site \textbf{r}, respectively, and $\omega_0$ is the phonon frequency. The sum only runs over those sites of the TI that are connected to a local phonon mode. To evaluate the fermionic self-energy ${\hat \Sigma}_{ph}$ arising from such an electron-phonon interaction, we consider a limit in which temperature is much larger than
the phonon frequency $\omega_0$ (i.e., the high-temperature approximation introduced in Ref. ~\cite{Bihary2005}). In this case, one retains only those terms in ${\hat \Sigma}_{ph}$ that contain the Bose distribution function since in this limit $n_B(\omega_0)\gg 1$.  The local fermionic self-energy at a site $\textbf{r}$ in the TI is computed self-consistently using the full Green's function of Eqs.(\ref{eq:fullGa}) and (\ref{eq:fullGb}), and given by
\begin{align}
\Sigma_{\textbf{r r}}^{r,<}(\omega) =  i g^{2} \int \frac{d\nu}{2\pi} D^{<}(\nu)
G_{\textbf{r r}}^{r,<}( \omega - \nu) \ ,
\label{eq:SE}
\end{align}
where
\begin{subequations}
\begin{align}
D_{0}^{<}(\omega) =& 2 i n_{B}(\omega) {\rm Im} D_{0}^{r} (\omega)  \\
D_{0}^{r}(\omega) =& \frac{1}{\omega - \omega_0 + i\delta} - \frac{1}{\omega + \omega_0 + i\delta}
\end{align}
\end{subequations}
are the lesser and retarded phonon Green's functions, which we assume to remain unchanged in
the presence of an applied bias.

To obtain an analytical expression for the lesser self-energy in Eq.(\ref{eq:SE}), we consider the limit
$\omega_{0} \rightarrow 0$ in which the self-energy, to leading order in
$k_{B}T/\omega_{0}$, becomes
\begin{align}
\Sigma_{{\bf r r}}^{r,<}(\omega) &=  2 g^{2} \frac{k_{B}T}{\omega_0} G_{{\bf r
r}}^{r,<}(\omega) \equiv \gamma G_{{\bf r r}}^{r,<}(\omega) \ .
\label{eq:sigma}
\end{align}
We next introduce the superoperator \cite{Bihary2005} $\tilde{D}$ which, when
operating on a Green's function matrix, returns the same matrix with all
elements set to zero except for those diagonal elements that represent sites at which an electron-phonon interaction exists, e.g.,
\begin{equation}
[\tilde{D} {\hat G}^{r,<}]_{{\bf r r^\prime}} = \left\{
\begin{array}{l}
G^{r,<}_{\bf r r^\prime} \delta_{\bf r,r^\prime} \text{ \ \ \ \ if
an electron-phonon interaction exists at ${\bf r}$} \\
0\qquad \qquad \text{ \ \ otherwise }%
\end{array}%
\right.
\end{equation}
and thus
\begin{equation}
\Sigma^{r,<}(\omega) = \gamma \tilde{D} {\hat G}^{r,<} \ .
\end{equation}
We next define the operator ${\hat U}$ that acts on a matrix ${\hat X}$ via
\begin{equation}
{\hat U}{\hat X} = {\hat G}^{r}{\hat X}{\hat G}^{a} \ .
\end{equation}
The solutions of the Dyson equations, Eqs.(\ref{eq:fullGa})and (\ref{eq:fullGb}), are then given by
\begin{subequations}
\begin{align}
\hat{G}^{<} &= \hat{U} \left[ 1 - \gamma {\tilde D} {\hat U} \right]^{-1} {\hat
\Lambda} \label{eq:Gless1} \\
\hat{G}^{r} &= \left[ 1 - \hat{g}^{r} \left( \hat{t} + \gamma {\tilde D} {\hat
G}^r \right) \right]^{-1} \hat{g}^{r} \ ,
\end{align}
\end{subequations}
where we defined the diagonal matrix $\hat{\Lambda} = \hat{g}_{r}^{-1}
\hat{g}^{<} \hat{g}_{a}^{-1}$. Note that the only non-zero elements of
$\hat{\Lambda}_{\bf r r}$ are those where ${\bf r}$ is a lead site. These
elements also contain the chemical potentials of the left and right leads. By
expanding the right hand side of Eq.(\ref{eq:Gless1}), we obtain
\begin{align}
\hat{G}^{<}_{\bf r r^\prime}  & =  \sum_{\bf l} {\hat G}_{\bf r l}^{r} \left
[{\hat \Lambda}_{\bf l l} + \gamma \sum_{\bf m} {\hat Q}_{\bf lm} {\hat
\Lambda}_{\bf m m}   +  \gamma^2 \sum_{\bf m,p} {\hat Q}_{\bf lm} {\hat Q}_{\bf mp}
{\hat \Lambda}_{\bf p p} + ... \right] {\hat G}_{\bf l r^\prime}^{a}  \ ,
\end{align}
where
\begin{equation}
{\hat Q}_{\bf lm}=\left\{
\begin{array}{l}
\left|G_{\bf lm}^{r}\right|^{2} \text{ \ \ \ \ if an electron-phonon interaction exists at ${\bf l}$}\\
0 \text{ \ \ \ \ \ \ \ \ \ \ \ \ otherwise}%
\end{array}%
\right.
\end{equation}
Defining next the vector ${\bm \lambda}$ with ${\bm \lambda}_{\bf m} = {\hat
\Lambda}_{\bf m m}$, we finally obtain
\begin{equation}
\hat{G}^{<}_{\bf r r^\prime}  =  \sum_{\bf l} {\hat G}_{\bf r l}^{r} \left [
\left(1 - \gamma {\hat Q} \right)^{-1} {\bm \lambda} \right]_{\bf l} {\hat
G}_{\bf l r^\prime}^{a}
\end{equation}
or $\hat{G}^{<} =  {\hat G}^{r} {\tilde \Sigma} {\hat G}^{a}$
where the diagonal matrix ${\tilde \Sigma}$ is defined via
\begin{equation}
{\tilde \Sigma}_{\bf ll}=  \left [ \left(1 - \gamma {\hat Q} \right)^{-1} {\bm
\lambda} \right]_{\bf l} \ .
\end{equation}

\subsection{Effect of Defects on the Excitation Spectrum of Edge States}
\label{sec:defect}

\subsubsection{Non-magnetic, elastically scattering defect}
\label{sec:TheoryPotDefect}

To understand the effects of a non-magnetic elastically scattering defect on the electronic spectrum of the edge states, we compute the full Green's functions in the presence of such a defect. The scattering off such a defect is described by the Hamiltonian in Eq.(\ref{eq:U0}). The full Green's function in Matsubara frequency space in the presence of a non-magnetic defect located at site ${\bf R}$ is then obtained from the perturbative expansion
\begin{align}
G({\bf r},{\bf r},\sigma ,i\omega _{n}) &= G_{0}({\bf r},{\bf r},\sigma ,i\omega
_{n})+G_{0}({\bf r},{\bf R},\sigma ,i\omega _{n}) U_0 G_{0}({\bf R},{\bf r},\sigma
,i\omega _{n}) \nonumber \\
&  \qquad +G_{0}({\bf r},{\bf R},\sigma ,i\omega _{n}) U_0  G_{0}({\bf R},{\bf R},\sigma
,i\omega _{n}) U_0 G_{0}({\bf R},{\bf r},\sigma ,i\omega _{n})+... \nonumber \\
& =G_{0}({\bf r},{\bf r},\sigma ,i\omega _{n})+\frac{G_{0}({\bf r},{\bf R},\sigma ,i\omega _{n}) U_0 G_{0}({\bf R},{\bf r},\sigma
,i\omega _{n})}{1-U_0 G_{0}({\bf R},{\bf R},\sigma,i\omega _{n})} \ ,
\end{align}
where $G_{0}({\bf r},{\bf r'}, \sigma ,i\omega_{n})$ is the non-local Matsubara Green's function of electrons with spin $\sigma$ of the clean TI.
If we set ${\bf r}={\bf R}$ we  have
\begin{equation}
G^{-1}({\bf R},{\bf R},\sigma ,i\omega _{n})=G_{0}^{-1}({\bf R},{\bf R},\sigma ,i\omega _{n})-U_0 \ .
\label{eq:fullG_U0}
\end{equation}
To obtain an analytic expression for the energies of the edge states in the presence of the defect, we consider a simplified model in which $G_{0}$
reflects the existence of only one set of a Kramers doublet of edge states, and is thus given by
\begin{equation}
G_{0}({\bf R},{\bf R},\sigma ,i\omega _{n})=\frac{Z_{\bf R}}{%
i\omega _{n}-E_{0}}+\frac{Z_{\bf R}}{i\omega _{n}+E_{0}}=\frac{2i Z_{\bf R} \omega _{n}}{\left(
i\omega _{n}\right) ^{2}-E_{0}^{2}}  \label{eq:G0}
\end{equation}
where $Z_{\bf R}$ is the spectral weight of the edge states at ${\bf R}$. We then obtain from Eq.(\ref{eq:fullG_U0})
\begin{equation}
G({\bf R},{\bf R},\sigma ,i\omega _{n})= \left[ \frac{Z_{\bf R}^+ }{i\omega_n - E_+} +  \frac{Z_{\bf R}^- }{i\omega_n - E_-} \right]
\end{equation}
with spectral weight
\begin{equation}
Z_{\bf R}^\pm = 1 \pm \frac{Z_{\bf R} U_0}{\sqrt{\left(Z_{\bf R} U_0 \right)^2+E_0^2}}
\end{equation}
and energies
\begin{equation}
E_\pm = \pm \sqrt{\left(Z_{\bf R} U_0 \right)^2+E_0^2} + Z_{\bf R} U_0
\end{equation}
Thus we find that, in the limit $Z_{\bf R} U_0 \ll E_0$, the energies to leading order in $Z_{\bf R} U_0$ are given by
\begin{equation}
E_\pm = \pm E_0 + Z_{\bf R} U_0
\end{equation}
reflecting a uniform, spin-independent shift of the edge states' energies.

\subsubsection{Magnetic Defect with Ising Symmetry}
\label{sec:Ising}

From Eq.(\ref{eq:Jscatt}), we have that the scattering Hamiltonian of a single magnetic defect with Ising-symmetry located at {\bf R} is given by
\begin{equation}
H_{Ising} =  J_z S \left( c^\dagger_{ {\bf R},\uparrow} c_{ {\bf R},\uparrow} - c^\dagger_{ {\bf R},\downarrow} c_{ {\bf R},\downarrow} \right)  \equiv \sum_\sigma {\bar J}_z^\sigma c_{{\bf R},\sigma }^{\dagger }c_{{\bf R},\sigma}
\end{equation}
where ${\bar J}_z^\sigma = J_z S \, {\rm sgn} \sigma$.
We can again compute the full Green's function in Matsubara frequency space from the perturbative expansion
\begin{align}
G({\bf r},{\bf r},\sigma ,i\omega _{n}) &= G_{0}({\bf r},{\bf r},\sigma ,i\omega
_{n})+G_{0}({\bf r},{\bf R},\sigma ,i\omega _{n}){\bar J}_{z}^\sigma G_{0}({\bf R},{\bf r},\sigma
,i\omega _{n}) \nonumber \\
&  \qquad +G_{0}({\bf r},{\bf R},\sigma ,i\omega _{n}) {\bar J}_{z }^\sigma  G_{0}({\bf R},{\bf R},\sigma
,i\omega _{n}) {\bar J}_{z }^\sigma G_{0}({\bf R},{\bf r},\sigma ,i\omega _{n})+... \nonumber \\
& =G_{0}({\bf r},{\bf r},\sigma ,i\omega _{n})+\frac{G_{0}({\bf r},{\bf R},\sigma ,i\omega _{n}){\bar J}_{z}^\sigma G_{0}({\bf R},{\bf r},\sigma
,i\omega _{n})}{1-{\bar J}_{z }^\sigma G_{0}({\bf R},{\bf R},\sigma,i\omega _{n})} \ ,
\end{align}
where $G_{0}({\bf r},{\bf r'}, \sigma ,i\omega_{n})$ is the non-local Matsubara Green's function of electrons with spin $\sigma$ of the clean TI. For ${\bf r}={\bf R}$ we  have
\begin{equation}
G^{-1}({\bf R},{\bf R},\sigma ,i\omega _{n})=G_{0}^{-1}({\bf R},{\bf R},\sigma ,i\omega _{n})-{\bar J}_{z
}^\sigma \ .
\label{eq:fullG_Jz}
\end{equation}
To obtain an analytic expression for the energies of the edge states in the presence of the defect, we again assume that $G_{0}$
reflects the existence of only one set of a Kramers doublet of edge states [see Eq.(\ref{eq:G0})], such that we obtain from Eq.(\ref{eq:fullG_Jz})
\begin{equation}
G({\bf R},{\bf R},\sigma ,i\omega _{n})= \left[ \frac{Z_{\bf R}^+ }{i\omega_n - E_+} +  \frac{Z_{\bf R}^- }{i\omega_n - E_-} \right]
\end{equation}
with spectral weight
\begin{equation}
Z_{\bf R}^\pm = 1 \pm \frac{Z_{\bf R} \bar{J}_z^\sigma}{\sqrt{\left(Z_{\bf R}\bar{J}_z^\sigma\right)^2+E_0^2}}
\end{equation}
and energies
\begin{equation}
E_\pm = \pm \sqrt{\left(Z_{\bf R} \bar{J}_z^\sigma\right)^2+E_0^2} + Z_{\bf R} \bar{J}_z^\sigma
\end{equation}
Thus we find that, in the limit $Z_{\bf R} \bar{J}_z^\sigma \ll E_0$, the energies to leading order in $Z_{\bf R} \bar{J}_z^\sigma$ are given by
\begin{equation}
E_\pm = \pm E_0 + Z_{\bf R} \bar{J}_z^\sigma
\end{equation}
reflecting a uniform but spin-dependent shift of the edge states' energies.

\subsubsection{Magnetic Defect with $xy$-Symmetry}
\label{sec:JxyDefect}

From Eq.(\ref{eq:Jscatt}) we have that the scattering Hamiltonian of a magnetic defect with $xy$-symmetry located at site ${\bf R}$ is given by
\begin{equation}
H_{xy}={\bar J}_+ c^\dagger_{ {\bf R},\downarrow} c_{ {\bf R},\uparrow} + {\bar J}_{-} c_{{\bf R},\uparrow }^{\dagger }c_{{\bf R},\downarrow }
\end{equation}
where ${\bar J}_+ = J_\pm S (1+i)$, and ${\bar J}_- = J_\pm S (1-i)$. The  full Green's function is then given by
\begin{align}
G({\bf r},{\bf r},\sigma ,i\omega _{n}) &= G_{0}({\bf r},{\bf r},\sigma ,i\omega
_{n})+G_{0}({\bf r},{\bf R},\sigma ,i\omega _{n}) {\bar J}_{\pm }
G_{0}({\bf R},{\bf R},-\sigma ,i\omega _{n}) {\bar J}_{\mp } G_{0}({\bf R},{\bf r},\sigma
,i\omega _{n}) + ... \nonumber \\
&=G_{0}({\bf r},{\bf r},\sigma ,i\omega _{n})+\frac{G_{0}({\bf r},{\bf R},\sigma ,i\omega _{n}) {\bar J}_{\pm}
G_{0}({\bf R},{\bf R},-\sigma ,i\omega _{n}){\bar J}_{\mp }G_{0}({\bf R},{\bf r},\sigma
,i\omega _{n})}{1-  {\bar J}_{+ } {\bar J}_{- } G_{0}({\bf R},{\bf R},\sigma
,i\omega _{n})G_{0}({\bf R},{\bf R},-\sigma ,i\omega _{n})}
\end{align}
where as above $G_{0}$ is the Matsubara Green's function of the unperturbed TI. Note that ${\bar J}_{+ } {\bar J}_{- } = \left| {\bar J}_{+ } \right|^2 $. If we set ${\bf r}={\bf R}$ we have
\begin{equation}
G^{-1}({\bf R},{\bf R},\sigma ,i\omega _{n})=G_{0}^{-1}({\bf R},{\bf R},\sigma ,i\omega _{n})-\left| {\bar J}_{+ } \right|^2 G_{0}({\bf R},{\bf R},-\sigma ,i\omega _{n}) \ .
\end{equation}
To obtain an analytic expression for the energies of the edge states in the presence of the defect, we again use the simplified model for $G_{0}$ as above [see Eq.(\ref{eq:G0})]
which yields
\begin{equation}
G^{-1}({\bf R},{\bf R},\sigma ,i\omega _{n})=\frac{\left( i\omega _{n}\right)
^{2}-E_{0}^{2}}{2i Z_{\bf R} \omega _{n}}- \left| {\bar J}_{+ } \right|^2 \frac{2i Z_{\bf R} \omega _{n}}{\left(
i\omega _{n}\right) ^{2}-E_{0}^{2}}
\end{equation}
and thus
\begin{eqnarray}
G({\bf R},{\bf R},\sigma ,i\omega _{n})
&=&\frac{Z^+_{\bf R} }{i\omega _{n}-E_{++}}+\frac{Z^-_{\bf R}}{i\omega _{n}-E_{+-}}+\frac{Z^-_{\bf R}}{i\omega _{n}-E_{-+}}+\frac{Z^+_{\bf R}}{i\omega
_{n} - E_{--}} \ .
\end{eqnarray}
Here, the energies of the edge states are given by
\begin{subequations}
\begin{align}
E_{+\pm} &= \sqrt{E_{0}^{2}+ Z^2_{\bf R} \left| {\bar J}_{+ } \right|^2 } \pm  Z_{\bf R}  \left| {\bar J}_{+ } \right| \\
E_{-\pm} &= -\sqrt{E_{0}^{2} + Z^2_{\bf R} \left| {\bar J}_{+ } \right|^2} \pm  Z_{\bf R} \left| {\bar J}_{+ } \right|
\end{align}
\end{subequations}
and one has for the spectral weight
\begin{equation}
Z^\pm_{\bf R} = \frac{Z_{\bf R}}{2} \left[ 1 \pm \frac{ Z_{\bf R}\left| {\bar J}_{+ } \right|}{\sqrt{%
E_{0}^{2}+ Z^2_{\bf R} \left| {\bar J}_{+ } \right|^2}}\right] \ .
\end{equation}
In the limit $Z_{\bf R}  \left| {\bar J}_{+ } \right| \ll E_0$ we obtain to leading order
\begin{subequations}
\begin{align}
E_{+\pm} &= E_{0} \pm Z_{\bf R}  \left| {\bar J}_{+ } \right|  \\
E_{-\pm} &= -E_{0} \pm Z_{\bf R}  \left| {\bar J}_{+ } \right| \ .
\end{align}
\end{subequations}
 Thus, the coupling of the Kramers doublet edge states by a spin-flip scattering defect leads to an energy splitting of $2 Z_{\bf R}  \left| {\bar J}_{+ } \right|$.

\section{Evolution of Spatial Current Patterns from Edge to Bulk States}
\label{sec:evol}

\begin{figure}[t]
 \begin{center}
\includegraphics[width=9.5cm]{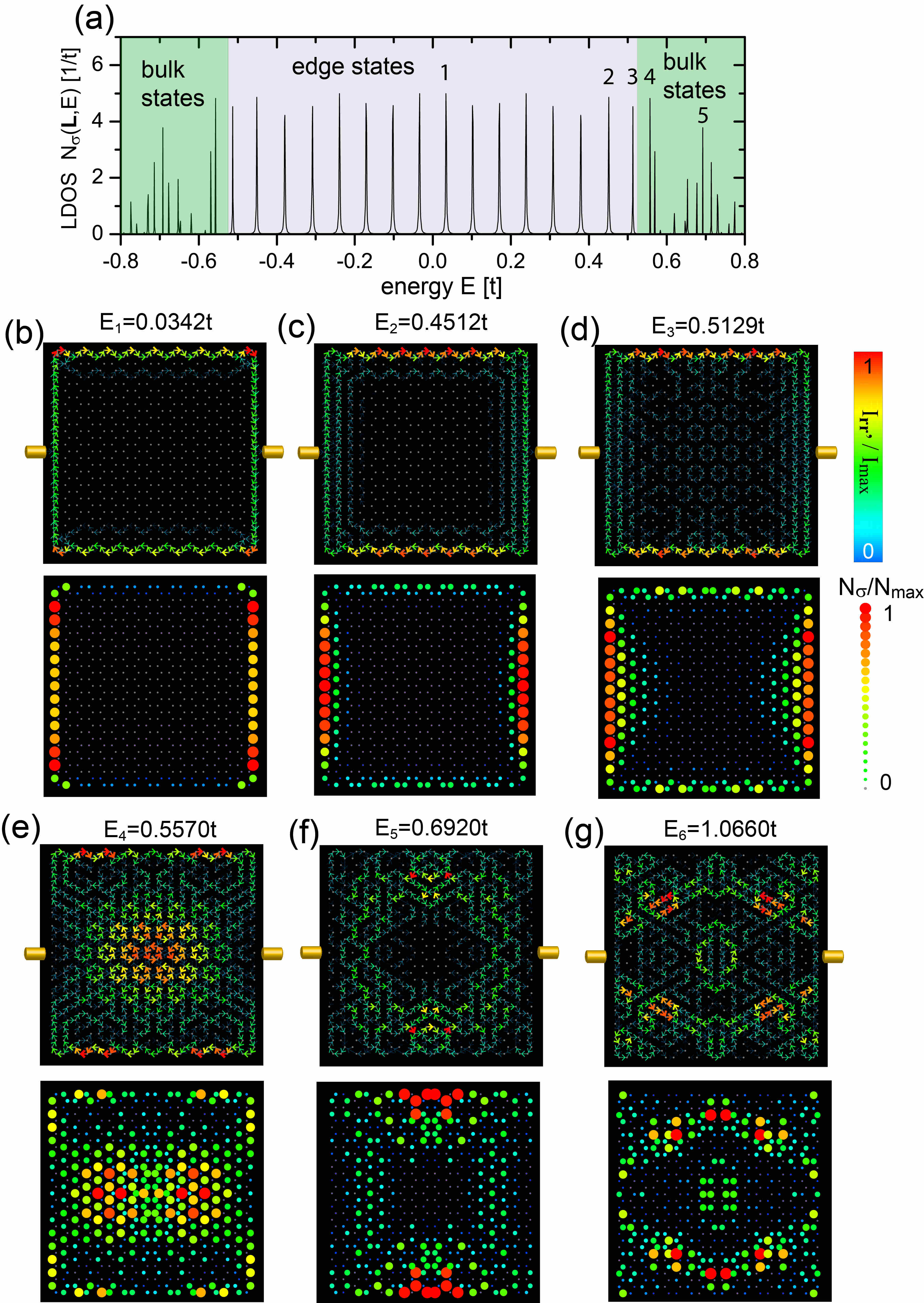}
\caption{(a) $N_\sigma(E)$ at \textbf{L} (see Fig.~\ref{fig:TI_lattice}) for a clean TI with $t_{l}=0.1t$. (b)-(d) Spatial patterns of $I^\uparrow_{\vect{r} \vect{r}'}$ (upper panels) and spatially resolved $N_\sigma$ (lower panels) for the edge states at $E_1 - E_3$ (see arrows 1-3 in (a)) and $V_g = E_i/e$. (e)-(g) $I^\uparrow_{\vect{r} \vect{r}'}$, (upper panels) and $N_\sigma$ (lower panels) for the bulk states at $E_4$, $E_5$ (see arrows 4 and 5 in (a)), and $E_6$ respectively. Color (see legends) and thickness of the arrows and dots represent the magnitude of the normalized current $I^\sigma_{\vect{r} \vect{r}'}/I^\sigma_{max}$ and the normalized local density of states $N_\sigma / N_{max}$, respectively (unless otherwise noted, normalization occurs for each panel separately in all figures).} \label{fig:DOS_currents}
\end{center}
\end{figure}

The spatial patterns of the local density of states and the spin-resolved charge currents associated with the edge states are qualitatively different from those of bulk states, and exhibit a characteristic evolution with increasing energy. To demonstrate this, we consider a two-dimensional TI with $N_a=9$ and $N_z=15$ \cite{VanDyke2016} (see Fig.~\ref{fig:TI_lattice})  and present
in Fig.~\ref{fig:DOS_currents}(a) its energy-dependent local density of states, $N_\sigma$, at \textbf{L}, for weak coupling to the leads, $t_{l}=0.1t$ (unless otherwise noted, all results shown in the following were obtained with $\Delta V=0.01t/e$, $k_B T=10^{-7}t$ and $\Lambda_R=0$).

In general, the TI - lead coupling gives rise to an energy broadening of both the edge and bulk states. However, the energy width of the edge states located below the spin-orbit gap $\Delta_{SO}=3\sqrt{3} \Lambda_{SO}$, i.e., at $|E| < \Delta_{so}$ (purple shaded background) is in general larger than that of the bulk states located at $|E| > \Delta_{so}$ (green shaded background).
This difference arises from the fact that the edge states (due to their localized nature) possess a larger spectral weight, $Z_\textbf{L}$, at \textbf{L} than the bulk states, as follows from a comparison of the local densities of states for an edge state shown in Fig.~\ref{fig:DOS_currents}(b) and for a bulk state shown in Fig.~\ref{fig:DOS_currents}(e). This in turn leads to a larger effective coupling, $Z_\textbf{L} t_{l}$, to the leads, and hence to a larger dephasing and energy width.

\begin{figure}[t]
 \begin{center}
\includegraphics[width=11.cm]{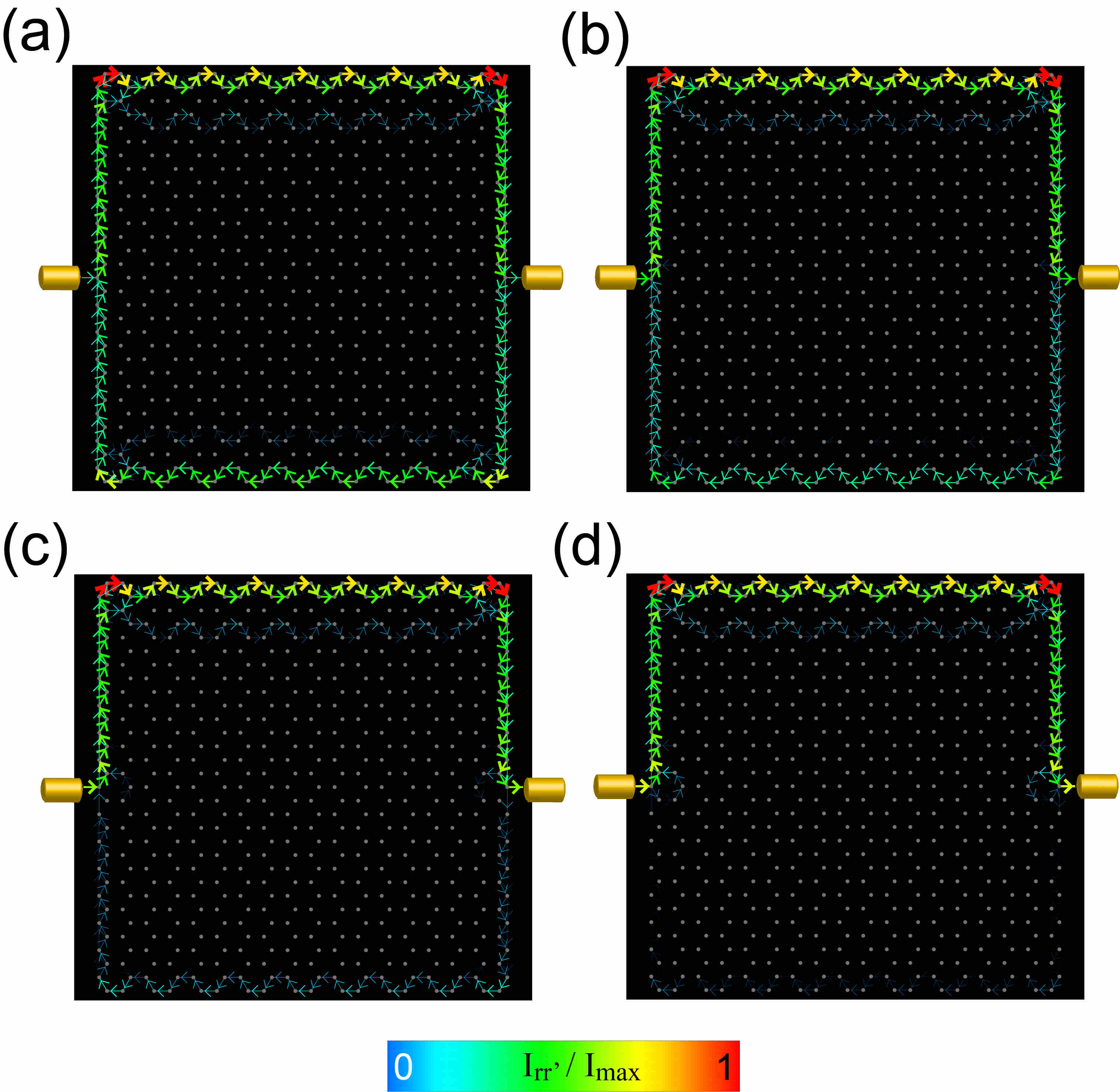}
\caption{$I^\uparrow_{\vect{r} \vect{r}'}$ for the edge state at $E_1=0.0342$t [Fig.~\ref{fig:DOS_currents}(b)] with increasing $t_l$: (a) $t_l=0.2$t, (b) $t_l=0.3$t, (c) $t_l=0.4$t, and (d) $t_l=0.5$t.}
\label{fig:current_TRL}
\end{center}
\end{figure}

In Fig.~\ref{fig:DOS_currents}(b) - (g) we present the evolution of the spatial patterns of the \su{} current, $I^\uparrow_{\vect{r} \vect{r}'}$, and the local density of states, $N_\sigma$, with increasing gate voltage $V_g$ and hence energy $E=eV_g$ of the states that carry the current, respectively. The spatial current patterns (upper panels) and local densities of states (lower panels) for the edge states located below the spin-orbit gap shown in Figs.\ref{fig:DOS_currents}(b), (c), and (d), corresponding to peaks 1, 2 and 3 in Fig.\ref{fig:DOS_currents}(a), respectively, are confined to the edges of the TI, and show an anisotropic spatial evolution into the bulk with increasing energy. This evolution reflects the anisotropy of the decay length of the edge states which increases with energy along the zig-zag edge but remains constant along the armchair edge \cite{Prada2013,Cano-Cortes2013}. As a result, the edge state wave functions and hence the corresponding local density of states, as well as the spatial current patterns, extend further into the bulk from the zig-zag edges as the states' energy approaches the spin-orbit gap. In contrast, for energies above the spin-orbit gap, the spatial current patterns (upper panels) and local densities of states (lower panels) shown in Figs.~\ref{fig:DOS_currents}(e), (f), and (g)  (with Figs.~\ref{fig:DOS_currents}(e) and (f) corresponding to peaks 4 and 5 in Fig.~\ref{fig:DOS_currents}(a)) reflect the nature of delocalized bulk states. Note the qualitative change in the current pattern and the local density of states between the highest energy edge state shown in Fig.~\ref{fig:DOS_currents}(d) and the lowest energy bulk state in Fig.~\ref{fig:DOS_currents}(e).

In Fig.~\ref{fig:current_TRL} we present the evolution of the current pattern of the lowest energy mode at $E_1=0.034$t [see Fig.~\ref{fig:DOS_currents}(b)] with increasing coupling, $t_l$, to the leads. While for weak coupling [Fig.~\ref{fig:DOS_currents}(b)], there is a substantial backflow of the \su-current along the lower side of the TI, this contribution decreases with increasing $t_l$ [Figs.~\ref{fig:current_TRL}(a)-(c)], until it is nearly completely suppressed for $t_l=0.5$t [Fig.~\ref{fig:current_TRL}(d)].
This also implies that while for small $t_l$, the net current flowing through the TI is substantially smaller than the largest current flowing inside the TI (the ratio is approximately $0.08$), the net current becomes comparable to the largest current inside the TI for larger $t_l$ (for $t_l=0.5$t, the ratio is approximately $0.68$).

\section{Effect of Time reversal symmetry preserving Defects on the Electronic Structure and Transport Properties of Topological Insulators}
\label{sec:potdefects}

We begin by considering the effects of defects that that do not break the time reversal symmetry of the TI, such as such as non-magnetic scatterers or  molecules with phonon modes. As we will show below, such defects, while not giving rise to backscattering or localization, nevertheless change the electronic spectrum of the TI and alter the flow of currents in their vicinity.

\begin{figure}[h]
 \begin{center}
\includegraphics[width=10cm]{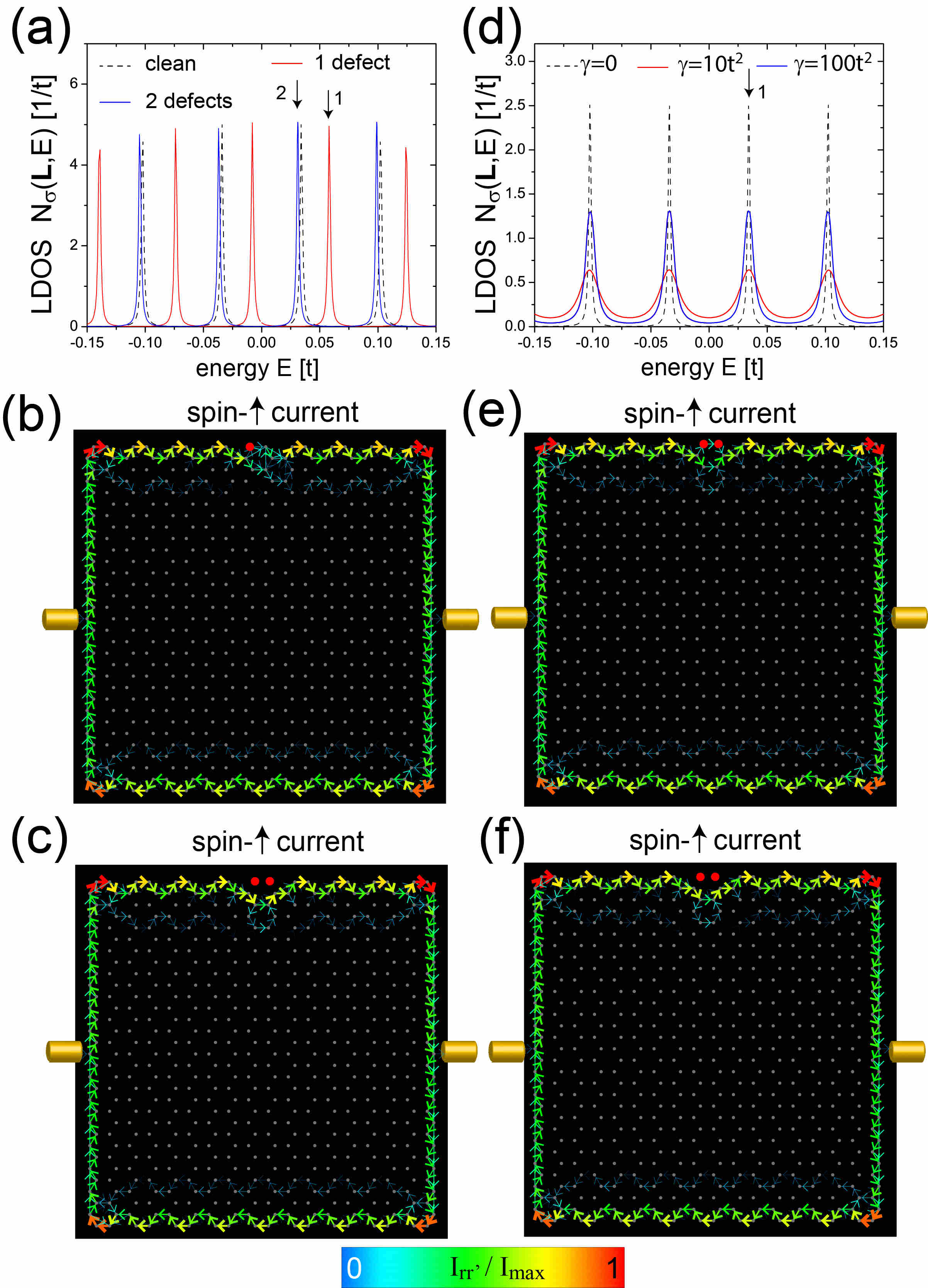}
\caption{(a) $N_\sigma({\bf L},E)$ for TIs with no, one, and two non-magnetic defects with scattering potential $U_0=10t$ and $t_{l}=0.1t$. (b) Spatial pattern of $I^\uparrow_{\vect{r} \vect{r}'}$ for a TI with a single defect (the location is indicated by a filled red circle) carried by the edge state at $E_1=0.0582t$ (arrow 1 in (a)). (c) $I^\uparrow_{\vect{r} \vect{r}'}$ with two defects carried by the edge state at $E_1=0.0312t$ (arrow 2 in (a)). (d)  $N_\sigma({\bf L}, E)$ for a TI coupled to two molecules with phonon modes (filled red circles) for different values of $\gamma$ [see Eq.(\ref{eq:sigma})]. Corresponding spatial patterns of $I^\uparrow_{\vect{r} \vect{r}'}$   for (e) $\gamma = 10t^2$, and (f) $\gamma = 100t^2$ carried by the lowest energy edge state at $E_1=0.034t$ (see arrow 1 in (d)).} \label{fig:potdefects}
\end{center}
\end{figure}

In Fig.~\ref{fig:potdefects}(a) we present the local density of states for a TI with one and two non-magnetic defects; their locations are indicated by filled red circles in Figs.~\ref{fig:potdefects}(b) and (c), respectively. The density of states shows that as expected, the particle-hole symmetry of the TI is broken by such defects, leading to a spin-independent shift of the excitation spectrum, as analytically shown in Sec.~\ref{sec:TheoryPotDefect}. We note that while the defect shifts the energy of the edge states, the differential conductance associated with charge transport through the edge states remains unaffected by the defects and is still given by the quantum of conductance, $G_0 = 2 e^2/\hbar$. Interestingly enough, the addition of a second defect can nearly reverse the effect of a single defect on the excitation spectrum, such that the excitation spectra of a clean system, and that containing two defects are nearly identical, as shown in Fig.~\ref{fig:potdefects}(a). This effect, of course, depends on the relative positions of the two defects. In Figs.~\ref{fig:potdefects}(b) and (c), we present the corresponding spatial patterns of the \su{} currents. While in the presence of a single defect, the current still flows through the defect site (Fig.~\ref{fig:potdefects}(b)), the current is nearly entirely expelled from the defect sites in the presence of two defects (Fig.~\ref{fig:potdefects}(c)). The latter result also explains our earlier observation: the combined scattering strength of the two defects is sufficient to essentially remove these two sites from the TI, thus (nearly) restoring the excitation spectrum of the clean TI. Similar effects of non-magnetic defects were recently also discussed in the context of the BHZ model of HgTe/CdTe quantum wells \cite{Sablikov2015}. Finally, we note that backscattering in the case of non-magnetic defects can only occur if it accompanied by simultaneous tunneling between opposite edges of the TI \cite{Lee2013}.

In contrast, the coupling of two TI sites to molecules with phonon modes leads to an energy broadening of edge states, as seen in $N_\sigma$ at \textbf{L} shown in  Fig.~\ref{fig:potdefects}(d). This broadening first increases with increasing $\gamma$, but begins to decrease again around $\gamma=10t^2$. At this point, the energy width of the local TI states that are directly coupled to the phonon modes is so large that the effective coupling to other TI sites (that contain no phonon mode) becomes weak, leading to a decreased dephasing, and hence narrower peaks in the local density of states at \textbf{L}, as shown in Fig.~\ref{fig:potdefects}(d). In contrast, the energy width of the states at the TI sites that are directly coupled to the phonons increases further with increasing $\gamma$. Concomitant with this evolution of the electronic structure, we find that the current flowing through the two sites that interact with the phonon modes decreases (Fig.~\ref{fig:potdefects}(e)), until it is nearly completely suppressed for $\gamma = 100t^2$ (Fig.~\ref{fig:potdefects}(f)). We note that the coupling to the phonon mode does not shift the energy of the edge states, but broadens their energy width.

\begin{figure}[h]
 \begin{center}
\includegraphics[width=8cm]{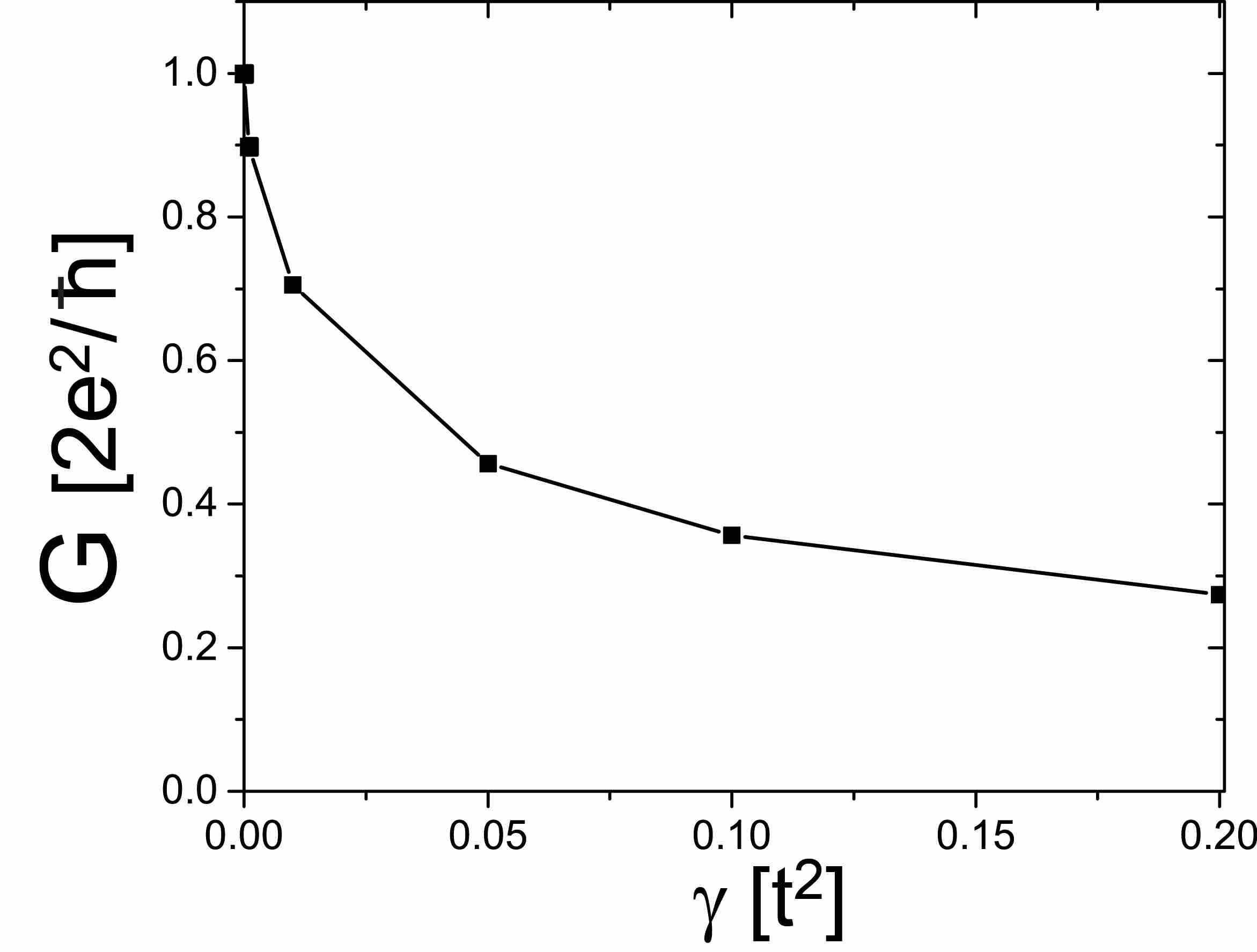}
\caption{Differential conductance of the lowest energy edge state at $E_1=0.034t$ (see Fig.~\ref{fig:potdefects}(d)) in the presence of two molecules with phonon modes as a function of $\gamma$.} \label{fig:conductance}
\end{center}
\end{figure}

While the interaction with phonon modes preserves the time-reversal symmetry of the TI, the induced broadening of the edge mode states leads to their hybridization with the bulk states, and hence starts to destroy the topological nature of the edge modes. We therefore expect that in contrast to elastically scattering non-magnetic defects which leave the differential conductance of the edge states unchanged, the presence of phonon modes (which scatter electrons inelastically), while not breaking the time-reversal symmetry of the TI, suppresses the conductance. To demonstrate this, we plot in Fig.~\ref{fig:conductance} the differential conductance of the TI in the presence of two molecules with phonon modes (see Fig.~\ref{fig:potdefects}(d)) as a function of $\gamma$.
Similar to normal metals \cite{Morr2016a}, we find that the presence of inelastically scattering phonon modes rapidly suppresses the conductance even of topologically protected edge states.

\section{Creation of Spin-Polarized Currents using Magnetic Defects}
\label{sec:magdefects}

As we discussed in Ref.\cite{VanDyke2016}, there are two different mechanisms by which highly spin-polarized currents can be created in 2D TI. The first mechanism is highly efficient in the limit where the energy width of the edge states is much smaller than their separation in energy (for example, due to a weak coupling to the leads, as shown in Fig.~\ref{fig:DOS_currents}(a), or a weak electron-phonon interaction). In this case, one can employ magnetic defects with Ising symmetry to lift the degeneracy of the Kramers pair of spin-$\uparrow$ and spin-$\downarrow$ bands by shifting their energies in opposite directions (see Sec.~\ref{sec:Ising}). The lifted degeneracy then allows one to select a non-degenerate spin-polarized state for current transport via gating, achieving high spin polarizations of the charge current up to 99\%. As one lifts the degeneracy of the Kramers pair of edge states, the differential conductance associated with charge transport through either of the states is now half the quantum of conductance.

\begin{figure}[h]
 \begin{center}
\includegraphics[width=10.cm]{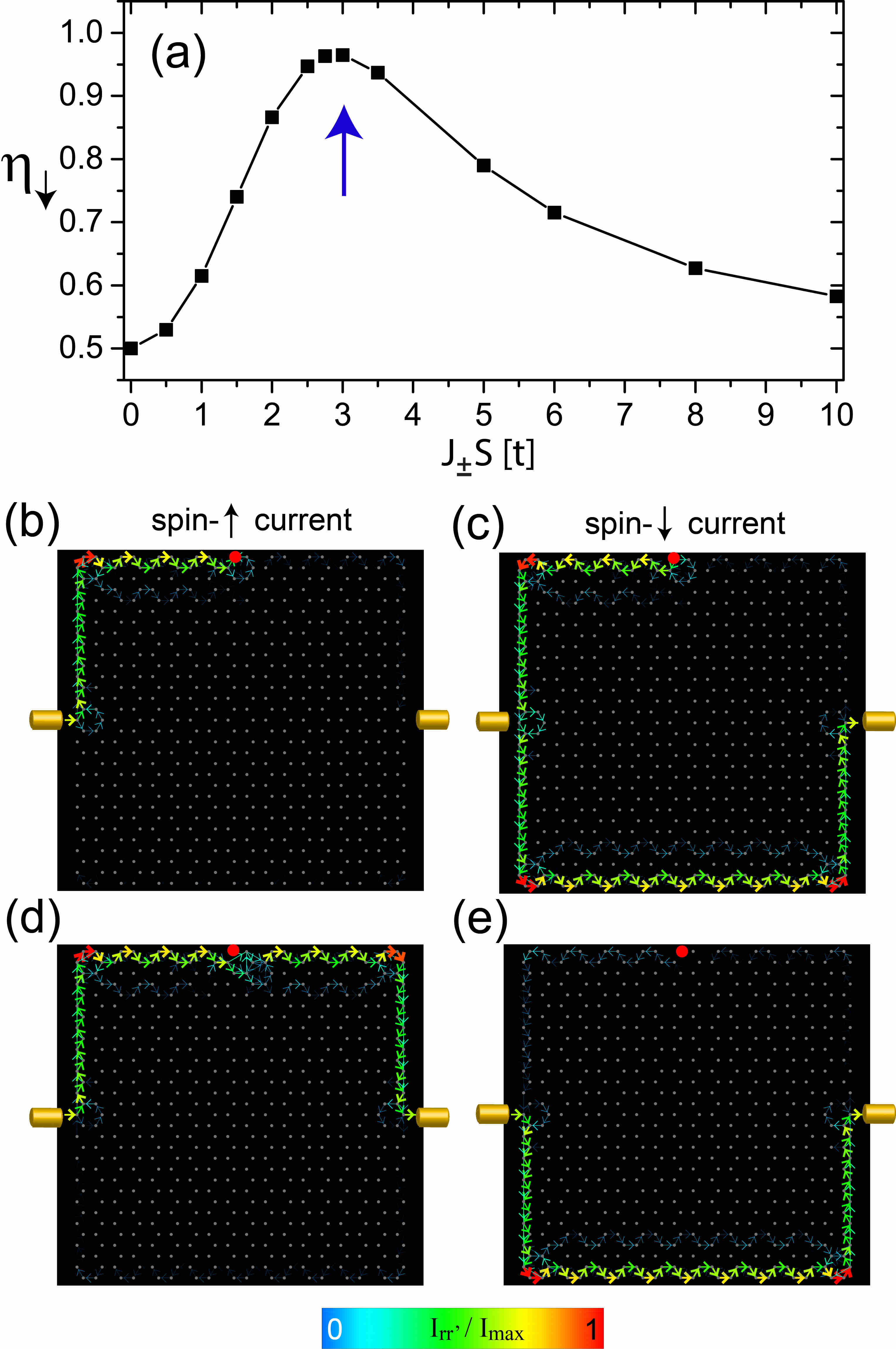}
\caption{(a) $\eta_\downarrow$ as a function of the scattering strength of a single magnetic (spin-flip) defect, $J_\pm S$ (see filled red circle in (b)). (b) $I^\uparrow_{\vect{r} \vect{r}'}$ and (c) $I^\downarrow_{\vect{r} \vect{r}'}$ at $J_\pm S=3$t where $\eta_\downarrow$ is maximal. (d) $I^\uparrow_{\vect{r} \vect{r}'}$ and (e) $I^\downarrow_{\vect{r} \vect{r}'}$ for $J_\pm S=20$t with $\eta_\downarrow=0.52$.}
\label{fig:1magdef}
\end{center}
\end{figure}

In contrast, when the edge states overlap in energy, for example due to a large coupling to the leads, magnetic (spin-flip) defects of $xy$-symmetry provide a qualitatively different, but equally efficient mechanism for the creation of spin-polarized currents. These defects scatter electrons between the spin-$\uparrow$ and spin-$\downarrow$ bands, leading to their hybridization (see Sec.~\ref{sec:JxyDefect}). In particular, for large couplings to the leads, the paths for the spin-$\uparrow$ and spin-$\downarrow$ currents are spatially well separated, as shown for the case of $t_l=0.5$t in Fig.\ref{fig:current_TRL}(d), such that when a spin-flip scattering defect is placed into the path of the spin-$\uparrow$ current [as shown in Fig.~\ref{fig:1magdef}(b)], they scatter nearly all of the current into the spin-$\downarrow$ band [as shown in Fig.~\ref{fig:1magdef}(c)]. This effectively blocks the spin-$\uparrow$ current and creates an additional contribution to the spin-$\downarrow$ current besides the one directly entering from the lead (a similar effect can occur in the chiral edge states of graphene \cite{Abanin2006}). As this scattering process blocks the current path for one of the two spin channels, it reduces the differential conductance of the TI to half of a quantum of conductance.
In Fig.~\ref{fig:1magdef}(a), we present the resulting spin polarization of the spin-$\downarrow$  current, $\eta_\downarrow$, as a function of the scattering strength of the magnetic (spin-flip) defect, $J_\pm S$. A maximum in the value of the spin polarization, $\eta_\downarrow = 0.965$ is achieved for $J_\pm S=3$t  with the resulting spatial current patterns of the spin-$\uparrow$ and spin-$\downarrow$ currents shown in Figs.~\ref{fig:1magdef}(b) and (c), respectively. With increasing scattering strength, the spin polarization decreases from its maximum value as the defect shifts the energy of the local site to higher energies, and the site of the defect becomes effectively decoupled from the rest of the TI. This weakens the scattering process between the two bands, hence reduces the spin polarization and leads to changes in the current patterns from the ones at maximum spin polarization. In particular, for $J_\pm S=20$t, the spin polarization has been reduced to $\eta_\downarrow = 0.52$, and the spin-$\uparrow$ current flows around the defect site [Fig.~\ref{fig:1magdef}(d)], implying that this site has been effectively decoupled from the TI, and that only a small part of the spin-$\uparrow$ current is scattered into the spin-$\downarrow$  band [Fig.~\ref{fig:1magdef}(e)].

\begin{figure}[h]
 \begin{center}
\includegraphics[width=11.cm]{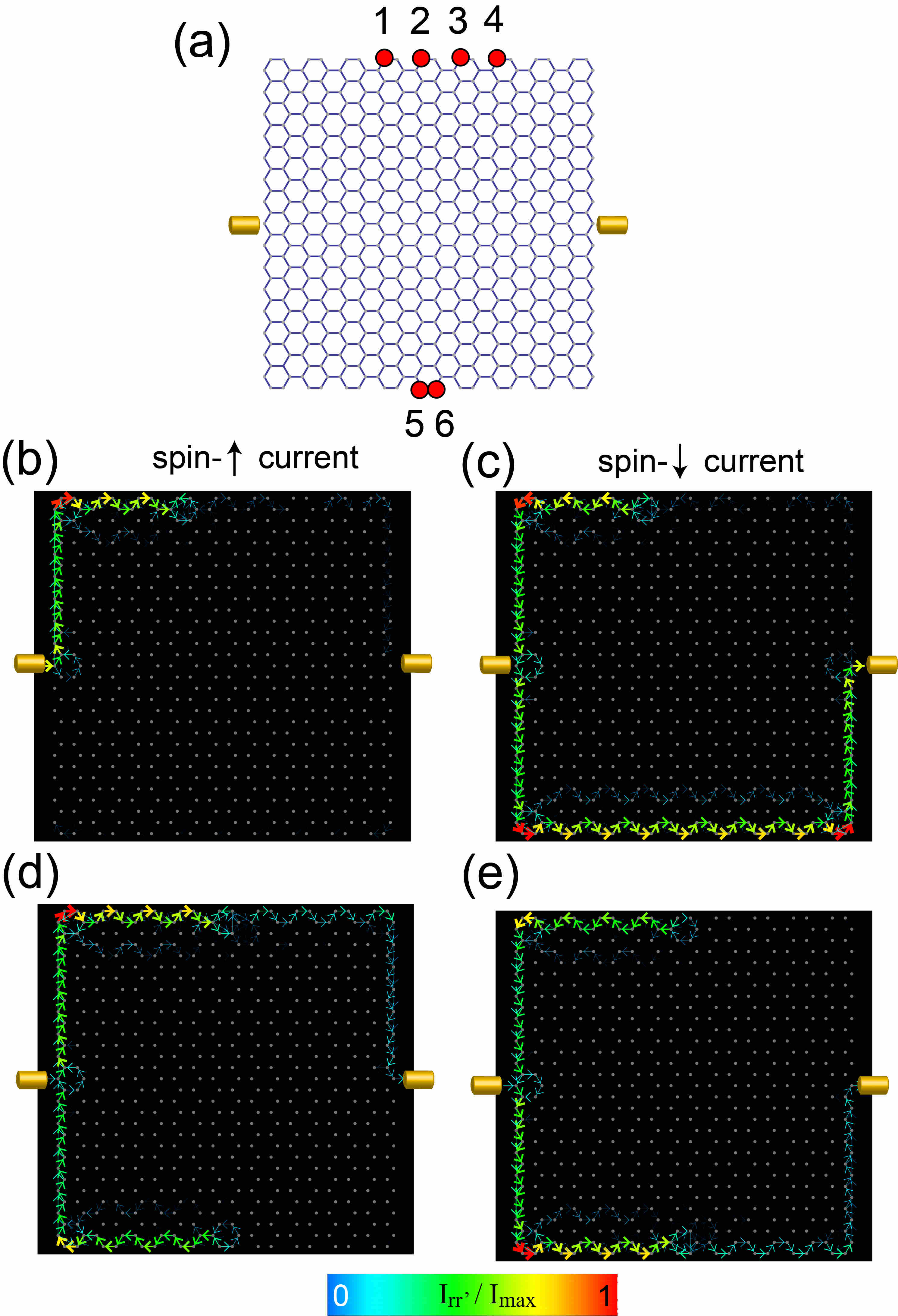}
\caption{(a) TI lattice and defect sites. Spatial current patterns for two defects with $J_\pm S=5$t and $t_l=0.5$t: (b) $I^\uparrow_{\vect{r} \vect{r}'}$, and (c) $I^\downarrow_{\vect{r} \vect{r}'}$ for two defects located at sites $1$ and $4$ in (a). (d) $I^\uparrow_{\vect{r} \vect{r}'}$, and (e) $I^\downarrow_{\vect{r} \vect{r}'}$ for two defects located at sites $2$ and $5$ in (a). }
\label{fig:2magdef}
\end{center}
\end{figure}

However, even in the case where the scattering strength of the magnetic defect exceeds the optimal value (i.e., the value where $\eta_\downarrow$ is largest), a high spin polarization can be achieved by adding additional defects. This is shown in Fig.~\ref{fig:2magdef} where we compare how different locations of two magnetic defects with $J_\pm S=5$t affect the spin polarization.
When the two defects are located at positions $1$ and $4$ [see Fig.~\ref{fig:2magdef}(a)] (giving rise to the current patterns shown in Figs.~\ref{fig:2magdef}(b) and (c)) or 2 and 3 (this case is considered in Fig. 2(d)--(f) of Ref.\cite{VanDyke2016}), the resulting spin polarizations are quite similar, with $\eta_\downarrow=0.95$ for the former case, and $\eta_\downarrow=0.965$ for the latter. Note that in both cases, the defects are located in the same sublattice of the TI, and that the resulting spin polarization exceeds that of a single defect with the same scattering strength, which when located at position 2 yields $\eta_\downarrow=0.79$ [see Fig.~\ref{fig:1magdef}(a)]. However, there are combination of sites for the two defects where the spin polarization is smaller than that of a single defect. The first case is that in which the two defects are located at symmetric positions, such as $2$ and $5$, with the resulting current patterns shown in Figs.~\ref{fig:2magdef}(d) and (e). Due to the symmetry of the positions, the current cannot be spin-polarized, and hence $\eta_\downarrow=\eta_\uparrow=0.5$. The same finding also holds when the defects are placed at the symmetric positions $2$ and $6$.

\begin{figure}[t]
 \begin{center}
\includegraphics[width=8.50cm]{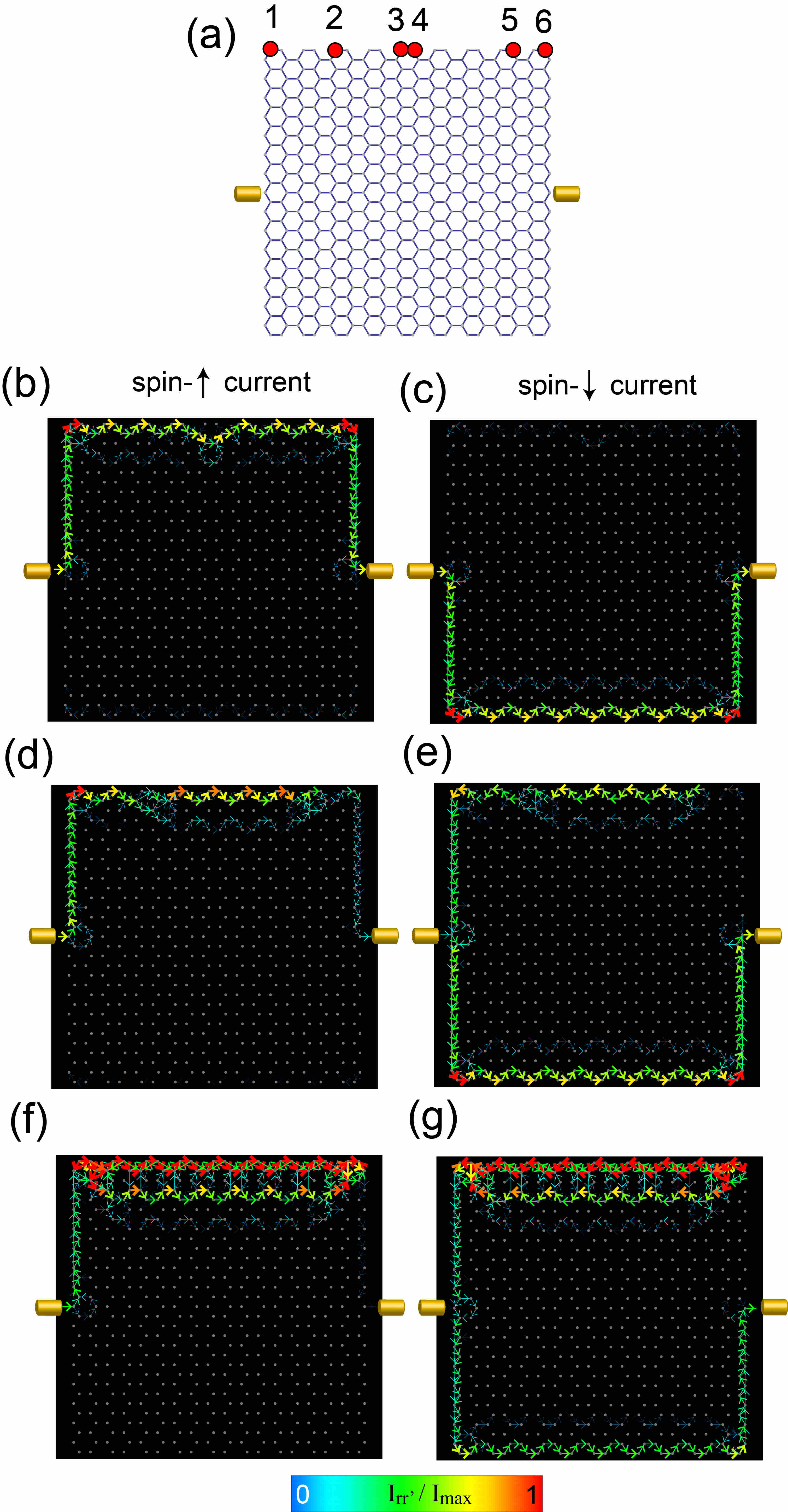}
\caption{(a) TI lattice and defect sites. Spatial current patterns for two defects with $J_\pm S=5$t and $t_l=0.5$t: (b) $I^\uparrow_{\vect{r} \vect{r}'}$, and (c) $I^\downarrow_{\vect{r} \vect{r}'}$ for two defects located at sites $3$ and $4$ in (a). (d) $I^\uparrow_{\vect{r} \vect{r}'}$, and (e) $I^\downarrow_{\vect{r} \vect{r}'}$ for two defects located at sites $2$ and $5$ in (a). (f) $I^\uparrow_{\vect{r} \vect{r}'}$, and (g) $I^\downarrow_{\vect{r} \vect{r}'}$ for two defects located at sites $1$ and $6$ in (a).}
\label{fig:2magdef_sub}
\end{center}
\end{figure}

The second case is when the defects are located in close proximity, and in different sublattices, as shown in Fig.~\ref{fig:2magdef_sub}.
When the two defects are located on neighboring sites [sites $3$ and $4$ in Fig.~\ref{fig:2magdef_sub}(a)], the combined scattering strenth is sufficiently large such that the sites
become effectively decoupled from the TI [similar to the case considered in Figs.~\ref{fig:1magdef}(d) and (e)] and the spin-$\uparrow$ current [Fig.~\ref{fig:2magdef_sub}(b)] flows around these two defect sites, with only minimal scattering into the spin-$\downarrow$ band [Fig.~\ref{fig:2magdef_sub}(c)]. As a result, the spin polarization is negligible $\eta_\downarrow \approx 0.5$.
When the two defects are located further apart [sites $2$ and $5$ in Fig.~\ref{fig:2magdef_sub}(a)], the spin polarization increases to $\eta_\downarrow=0.8$, concomitant with a significant scattering of the spin-$\uparrow$ current [Fig.~\ref{fig:2magdef_sub}(d)] into the spin-$\downarrow$ band [Fig.~\ref{fig:2magdef_sub}(e)]. Finally, when the two defects are located at the corner sites of the TI [sites $1$ and $6$ in Fig.~\ref{fig:2magdef_sub}(a)], the spin polarization increases to $\eta_\downarrow=0.9$. Interestingly enough, there is now a significant current in the spin-$\uparrow$ and spin-$\downarrow$ bands that is scattered back and forth between the two defect sites, as shown in Figs.~\ref{fig:2magdef_sub}(f) and (g), respectively.

\begin{figure}[h]
 \begin{center}
\includegraphics[width=11.cm]{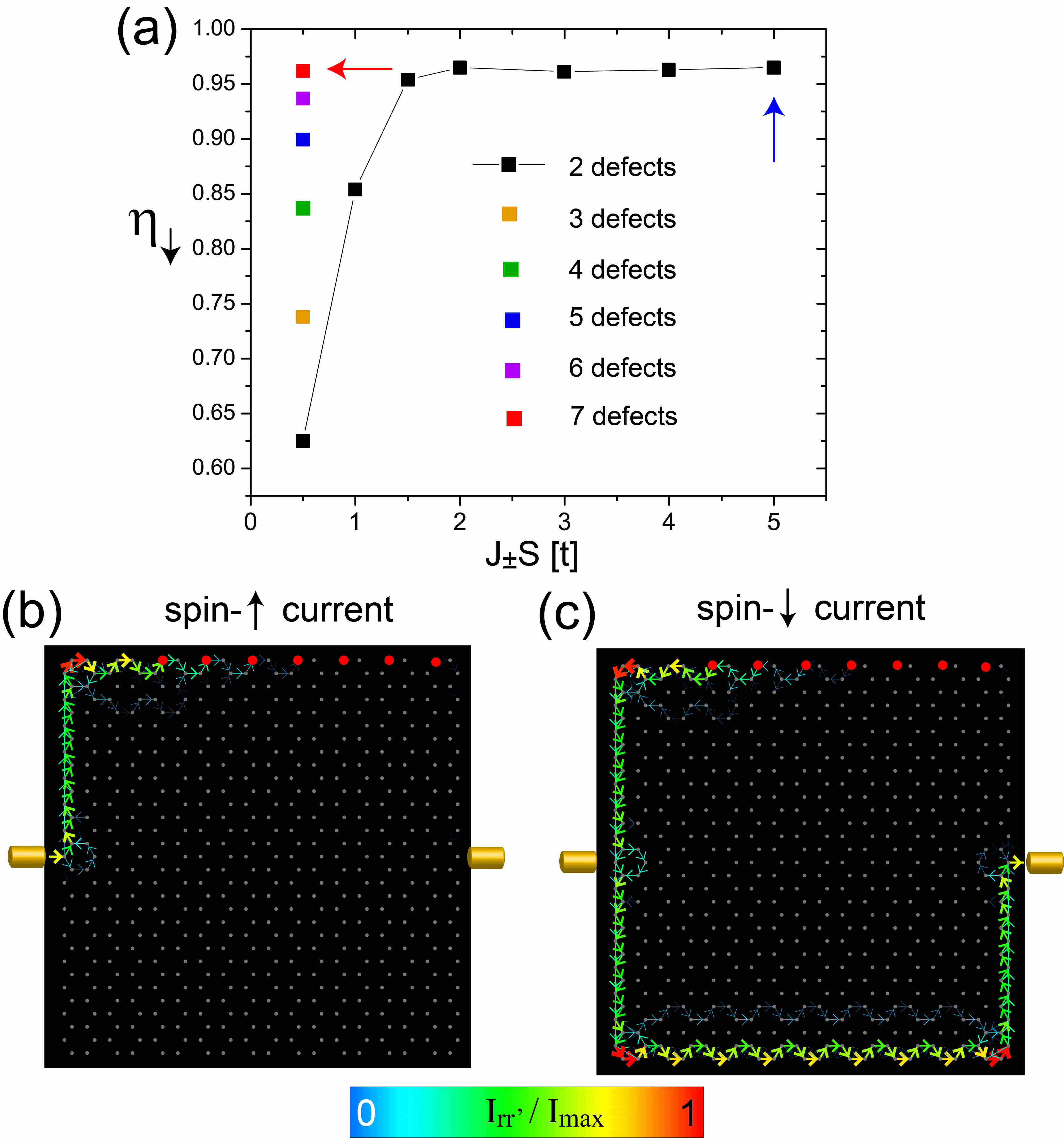}
\caption{(a) $\eta_\downarrow$ as a function of $J_\pm S$ for multiple defects and $t_{l}=0.5t$. (b) $I^\uparrow_{\vect{r} \vect{r}'}$, and (c) $I^\downarrow_{\vect{r} \vect{r}'}$ for $J_\pm S= 0.5t$, $V_g= 0.025t/e$ and 7 magnetic defects (see red arrow in (a)).}
\label{fig:multi_defects}
\end{center}
\end{figure}
 Similarly, when the scattering strength of defects is smaller than the optimal value, a high spin polarization can be achieved by adding additional defects. Consider for example the spin polarization, $\eta_\downarrow$, for two defects located at positions $2$ and $3$ (of Fig.~\ref{fig:2magdef}(a)) as a function of scattering strength, $J_\pm S$, with $t_{l}=0.5t$, shown in Fig.~\ref{fig:multi_defects}(a). We find that $\eta_\downarrow$ only begins to decrease from its value of $\eta_\downarrow = 0.965$ once $J_\pm S$ becomes smaller than approximately $1.5t$ (note in the limit $J_\pm \rightarrow 0$, one has $\eta_\downarrow \rightarrow 0.5$). However, even for a rather small value of $J_\pm S = 0.5t$, where in the presence of two defects one has $\eta_\downarrow = 0.625$, we find that $\eta_\downarrow$ can be rapidly increased by increasing the number of defects in the TI. In this case, a spin polarization of $90 \%$ is already reached for 5 defects, and a value of $\eta_\downarrow = 0.962$ (corresponding to the large $J_\pm S$ limit for 2 defects) is obtained for 7 defects. For the latter case (7 defects with $J_\pm S = 0.5t$ and $\eta_\downarrow = 0.962$), we plot in Figs.~\ref{fig:multi_defects}(b) and (c) the spatial \su{} current and \sd{} current patterns, respectively. These patterns are qualitatively similar to those in Figs.2(e),(f) of Ref. \cite{VanDyke2016} (with nearly identical values of $\eta_\downarrow$), the only difference being that the scattering of the \su{} electrons into the \sd{} band in Figs.~\ref{fig:multi_defects}(b) and (c) occurs over a larger spatial range.

Similarly, for magnetic defects with Ising symmetry, the spin polarization remains unaffected as long as the splitting between the \su{} and \sd{} states (which decreases with decreasing $J_z S$ ) is larger than $e\Delta V$ with $\Delta V$ being the applied voltage. For the case considered in Fig.~2(a),(b) of Ref. \cite{VanDyke2016} with a single defect of Ising symmetry, we find that even for $J_z S = t$, the current is still highly spin-polarized with $\eta_\downarrow=0.951$. Only for smaller values of $J_z S$ does the spin polarization significantly decrease. However, even for a weakly scattering defect with $J_z S = 0.5t$, one still obtains $\eta_\downarrow=0.788$. We can therefore conclude that the creation of highly spin-polarized currents in finite two-dimensional TIs is a universal phenomenon that is not tied to particular values of the magnetic scattering strength, but can be achieved over a wide range of $J_{z,\pm} S$.

In what follows, we will demonstrate that the creation of highly spin-polarized currents is robust against changes in the strength of the spin-orbit coupling (Sec.~\ref{sec:SO_coupling}), changes in the size of the TI (Sec.~\ref{sec:sizeTI}), the width of the attached leads (Sec.~\ref{sec:widthleads}), the inclusion of a Rashba spin-orbit interaction (Sec.~\ref{sec:SORashba}), and the inclusion of dephasing arising from an electron-phonon interaction (Sec.~\ref{sec:dephasing}).

\subsection{Strength of the Spin-Orbit Coupling}
\label{sec:SO_coupling}

\begin{figure}[h]
 \begin{center}
\includegraphics[width=12.cm]{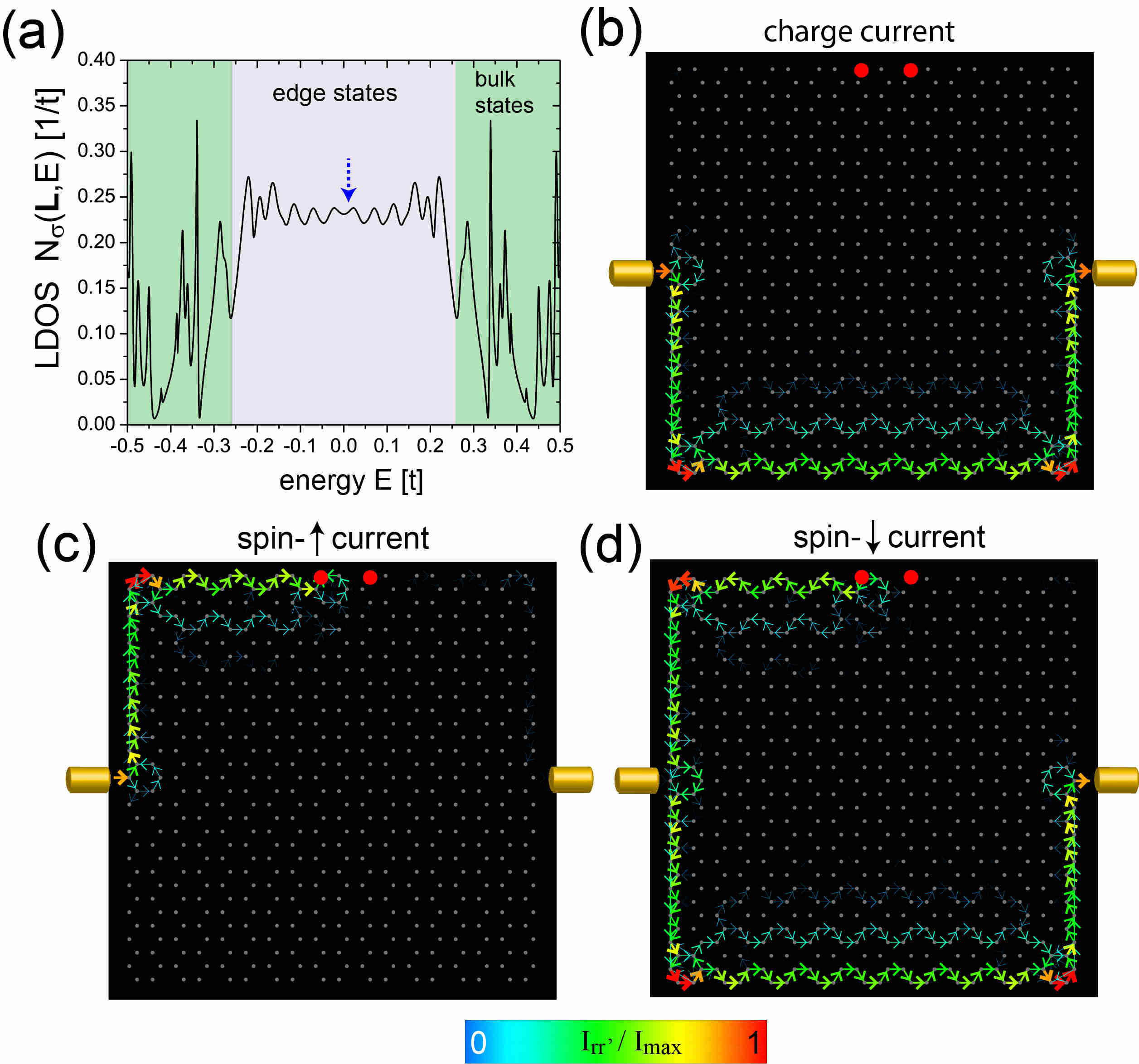}
\caption{(a) $N_\sigma({\bf L},E)$ for two magnetic defects of $xy$-symmetry (see filled red circles in (b)), $J_\pm S=5t$, $t_{l}=0.5t$, and $\Lambda_{SO}=0.05t$. (b) $I^c_{\vect{r} \vect{r}'}$, (c) $I^\uparrow_{\vect{r} \vect{r}'}$, and (d) $I^\downarrow_{\vect{r} \vect{r}'}$ for the edge state at $E_1 = 0.01t$ (see dashed blue arrow in (a)).}
\label{fig:LSO}
\end{center}
\end{figure}

One of the crucial elements in the creation of a topological insulator is the existence of the spin-orbit coupling with strength $\Lambda_{SO}$. As the topological edge states can only exist at energies smaller than the spin-orbit gap, $\Delta_{SO}$, the effect of creating spin-polarized currents is expected to diminish as $\Lambda_{SO} \rightarrow 0$. To investigate the effect of a reduced $\Lambda_{SO}$  on the magnitude of the currents' spin polarization, we consider in Fig.~\ref{fig:LSO} the case of two magnetic defects located at positions $2$ and $3$ of Fig.~\ref{fig:2magdef}(a) with $J_\pm S=5t$ and $t_{l}=0.5t$, but a value of $\Lambda_{SO}=0.05t$ which is half of that considered above.
The resulting LDOS, $N_\sigma({\bf L})$ shows as expected a smaller spin-orbit gap since $\Delta_{SO} = 3 \sqrt{3} \Lambda_{SO}$. However, the spatial patterns of the \su{} current (Fig.~\ref{fig:LSO}(c)) and the \sd{} current (Fig.~\ref{fig:LSO}(d)) are qualitatively similar to those shown above in Figs.~\ref{fig:1magdef}(b) and (c); the resulting charge current, $I^c_{\vect{r} \vect{r}'}$, is shown in Fig.~\ref{fig:LSO}(b). We note that the reduction in $\Lambda_{SO}$ leads to a larger decay length of the edge states, such that the currents shown in Figs.~\ref{fig:LSO}(b) - (d) extend further into the bulk than those shown in Figs.~\ref{fig:1magdef}(b) and (c). This leads to only a small reduction in the spin polarization with $\eta_\downarrow = 0.946$, while for the two-dimensional TI discussed in Ref. \cite{VanDyke2016} we obtained $\eta_\downarrow = 0.963$. These findings demonstrate that the ability to create spin-polarized currents is qualitatively and quantitatively robust against variations in the spin-orbit coupling. Only in the limit  $\Lambda_{SO} \rightarrow 0$ will the decrease in the size of the spin-orbit gap eventually lead to a vanishing of the current's spin polarization.

\subsection{Size and Aspect Ratio of the TI}
\label{sec:sizeTI}

\begin{figure}[t]
 \begin{center}
\includegraphics[width=11.cm]{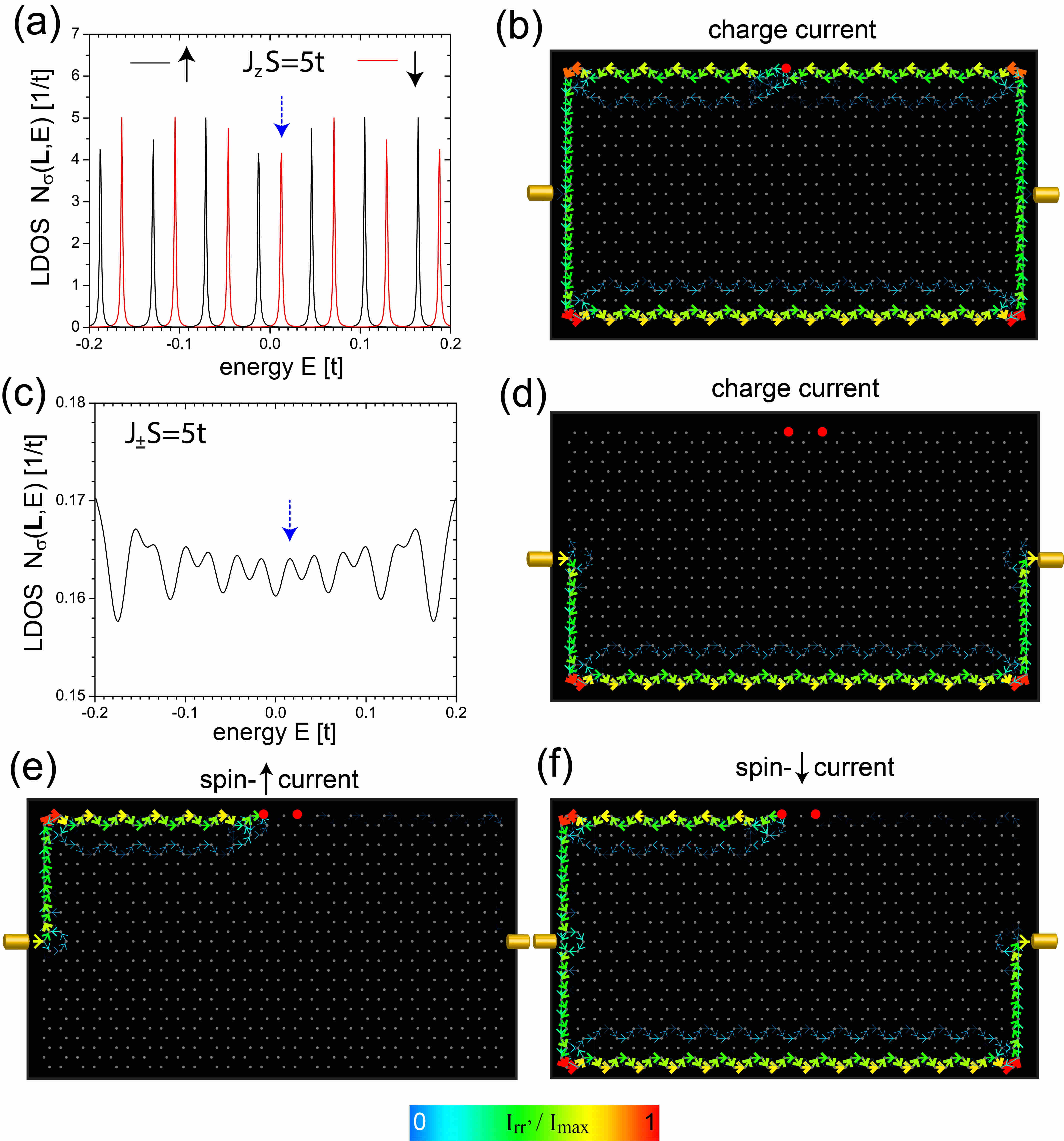}
\caption{(a) $N_\sigma({\bf L},E)$ in the presence of a magnetic defect (see filled red circle in (b)) with Ising symmetry, $J_z S=5t,J_\pm S =0$, and $t_{l}=0.1t$. (b) $I^c_{\vect{r} \vect{r}'}$ carried by the lowest energy edge state at $E_1=0.0142t$ (see dashed blue arrow in (a)). (c) $N_\sigma({\bf L},E)$ for two magnetic defects (see filled red circles in (d)) with $xy$-symmetry, $J_\pm S=5t, J_z S=0$, and $t_{l}=0.5t$ . (d) $I^c_{\vect{r} \vect{r}'}$, (e) $I^\uparrow_{\vect{r} \vect{r}'}$, and (f) $I^\downarrow_{\vect{r} \vect{r}'}$, carried by the edge state at $E_1 = 0.0175t$ (see dashed blue arrow in (c)).}
\label{fig:TI2827}
\end{center}
\end{figure}
To investigate the effect of a TI's size and aspect ratio, $N_a/N_z$, on the ability to create spin-polarized currents, we consider charge transport in a TI with $N_a = 14$ and $N_z = 13$.  In Fig.\ref{fig:TI2827}(a) we present the local density of states at $\textbf{L}$ for a TI containing a single magnetic defect of Ising symmetry with $J_zS=5t$ and $t_{l}=0.1t$ (the defect's location is denoted by a filled red circle in Fig.~\ref{fig:TI2827}(b)). As expected, the degeneracy of the Kramers doublet is lifted by the defect, such that a single spin-polarized state can be selected for charge transport. By accessing the state at $E_i=0.0125t$ (see dashed arrow in Fig.~\ref{fig:TI2827}(a)), we find that the charge current is 98.9\% \sd{} polarized ($\eta_\downarrow = 0.989$), which is similar to the spin polarization $\eta_\downarrow = 0.99$ obtained in Ref. \cite{VanDyke2016} for a TI with $N_a = 9$ and $N_z = 15$. Moreover, in the limit of large coupling to the leads with $t_{l}=0.5t$ and two defects with $xy$-symmetry and $J_\pm S = 5t$ (see Figs.~\ref{fig:TI2827}(c) - (f)), the spin polarization of the charge current (Fig.~\ref{fig:TI2827}(d)) is $\eta_\downarrow = 0.962$, and thus again similar to the value obtained for the cases shown in Fig.~\ref{fig:2magdef} and discussed in Ref.~\cite{VanDyke2016}. These results demonstrate that the ability to create spin-polarized currents is thus independent of the particular size or aspect-ratio of the nanoscopic TI.\\

\subsection{Width of the attached leads}
\label{sec:widthleads}

\begin{figure}[t]
 \begin{center}
\includegraphics[width=11.cm]{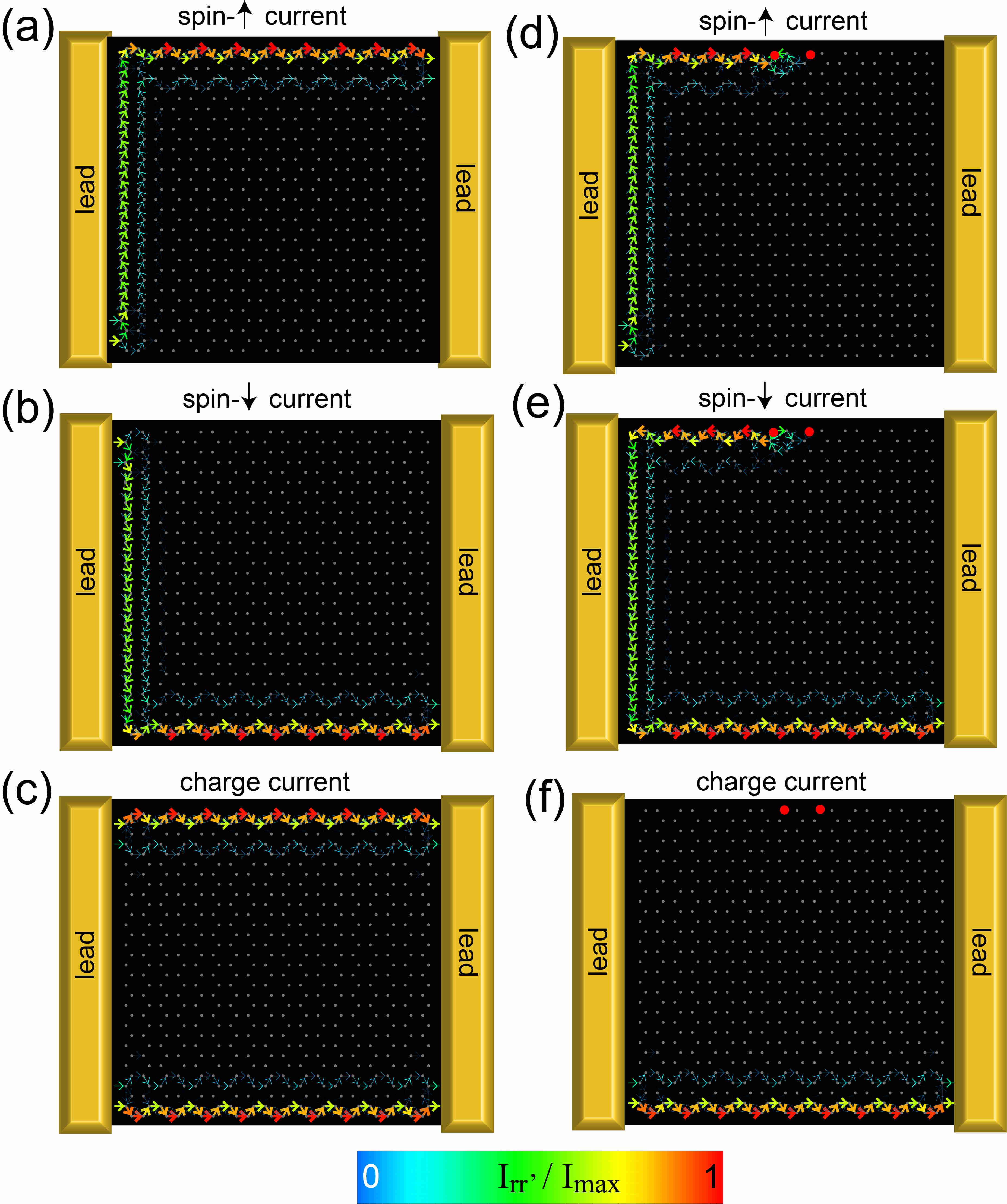}
\caption{(a) $I^\uparrow_{\vect{r} \vect{r}'}$, (b) $I^\downarrow_{\vect{r} \vect{r}'}$, and (c) $I^c_{\vect{r} \vect{r}'}$ in a TI attached to wide leads with $t_{l}=0.5t$ and $V_g = 0.1t/e$. (d) $I^\uparrow_{\vect{r} \vect{r}'}$, (e) $I^\downarrow_{\vect{r} \vect{r}'}$, and (f) $I^c_{\vect{r} \vect{r}'}$ for two magnetic defects (see filled red circles) with $xy$-symmetry and $J_{\pm} S= 5t$.} \label{fig:Fig9}
\end{center}
\end{figure}
The ability to create highly spin-polarized currents also remains unaffected by the width of the leads attached to the TI, as shown in Fig.~\ref{fig:Fig9} where we consider  the spatial current patterns in a TI attached to two wide leads.
In Figs.~\ref{fig:Fig9}(a)-(c), we present the \su{}, \sd{}, and charge currents through a clean TI (without any magnetic defects), respectively. It is interesting to note that there are only two sites on each side of the TI (for each spin degree) where an appreciable current enters or exits the TI.  In the presence of two magnetic defects with $xy$-symmetry and $J_\pm S=5t$, the \su{} current (see Fig.~\ref{fig:Fig9}(d)) is nearly completely scattered into a \sd{} current (Fig.~\ref{fig:Fig9}(e)), resulting in $\eta_{\downarrow}=0.996 $  and a charge current that flows nearly entirely along the lower edge of the TI (Fig.~\ref{fig:Fig9}(f)). In the presence of the defect, no \sd{} current enters the TI directly from the lead, being created instead by the scattering of the \su{} current by the magnetic defects.  This robustness of creating spin-polarized current in the presence of wide leads is encouraging for spintronics applications, since it obviates the need for atomically-sharp electrical contacts in circuit design.\\

\subsection{Rashba spin-orbit interaction}
\label{sec:SORashba}

The inclusion of a Rashba spin-orbit interaction leads to a spin-flip process when an electron hops between two nearest neighbor sites. As first shown by Kane and Mele \cite{Kane2005a} this can lead to the destruction of the topological insulator when the Rashba spin-orbit interaction becomes sufficiently large (i.e., for $\lambda_{R} > \lambda^c_{R} = 2 \sqrt{3} \lambda_{SO}$) and the bulk gap closes. The question therefore naturally arises to what extent the creation of spin-polarized currents is robust against the inclusion of a Rashba spin-orbit interaction.

\begin{figure}[h]
 \begin{center}
\includegraphics[width=8.5cm]{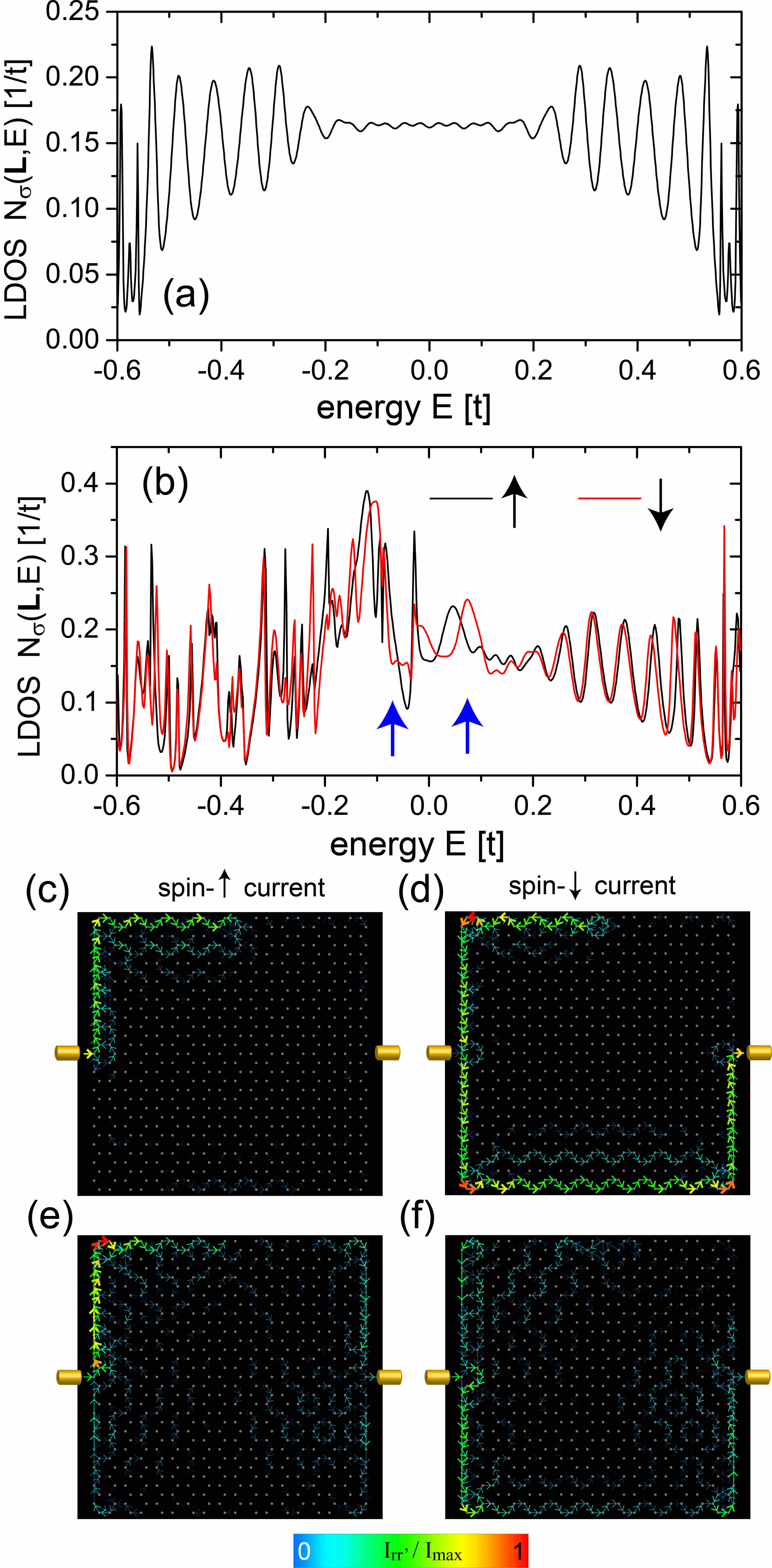}
\caption{Topological Insulator with $\Lambda_R=0.2t=2 \Lambda_{SO}$ and $t_{l}=0.5t$. $N_\sigma({\bf L},E)$ for (a) $\Lambda_R=0$ and (b) $\Lambda_R=0.2t$. (c) $I^\uparrow_{\vect{r} \vect{r}'}$ and (d)  $I^\downarrow_{\vect{r} \vect{r}'}$ for $V_g = +0.074t/e$. (e) $I^\uparrow_{\vect{r} \vect{r}'}$ and (f)  $I^\downarrow_{\vect{r} \vect{r}'}$ for $V_g = -0.074t/e$.}\label{fig:Rashba}
\end{center}
\end{figure}

In Figs.~\ref{fig:Rashba}(a) and (b), we compare the density of states, $N_\sigma({\bf L},E)$ for a TI with two spin-flip defects with $J_\pm S=5$t with zero Rashba interaction, Fig.~\ref{fig:Rashba}(a), and with $\Lambda_R=2 \Lambda_{SO}=0.2$t. As was shown previously for the case of a cylinder \cite{Laubach2014}, the inclusion of a Rashba spin-orbit interaction pushes the lower edge of the spin-orbit gap to higher energies. This is also confirmed by a comparison of $N_\sigma({\bf L},E)$ shown in Figs.~\ref{fig:Rashba}(a) and (b): the structure of $N_\sigma({\bf L},E)$ for $E\gtrsim 0$ and $\Lambda_R=2 \Lambda_{SO}$ still resembles approximately that of $N_\sigma({\bf L},E)$ for $\Lambda_R=0$, while the density of states are qualitatively different for $E<0$.
This conclusion is supported by the form of the spin polarization and current patterns carried by the states at energies $E=\pm 0.074$t, indicated by the two blue arrows in Fig.~\ref{fig:Rashba}(b). For $E=+ 0.074$t, the spatial patterns of $I^\uparrow_{\bf r,r^\prime}$ [Fig.~\ref{fig:Rashba}(c)] and $I^\downarrow_{\bf r,r^\prime}$ [Fig.~\ref{fig:Rashba}(d)] as well as a spin polarization of $\eta_\downarrow =0.973$ resemble those considered above in Fig.~\ref{fig:2magdef}, and hence imply that the charge transport is still carried by the topologically protected edge modes. In contrast, for $E= -0.074$t, the spatial patterns of $I^\uparrow_{\bf r,r^\prime}$ [Fig.~\ref{fig:Rashba}(e)] and $I^\downarrow_{\bf r,r^\prime}$ [Fig.~\ref{fig:Rashba}(f)] now exhibit a significant current density in the center of the TI, implying that bulk states are involved in charge transport. As a result, the spin polarization is significantly diminished to  $\eta_\downarrow =0.668$. Thus, our results demonstrate that even for a significant Rashba spin-orbit coupling of $\Lambda_R = 2 \Lambda_{SO}$, it is still possible to create highly spin-polarized currents. In general, the ability to create spin-polarized currents should only be lost once $\lambda_{R}$ exceeds  $\lambda^c_{R}$ since in this case, the bulk-gap is closed, and hence topologically protected edge states no longer exist.

\subsection{Dephasing arising from an electron-phonon interaction}
\label{sec:dephasing}

To investigate the effects of dephasing on the ability to create spin-polarized currents in TIs, we introduce a local electron-phonon interaction [see Eq.(\ref{eq:eph})] at each site of the TI, and compute the spin-resolved charge currents using the high-temperature approximation outlined in Sec.\ref{sec:theoryI}.

\begin{figure}[h]
 \begin{center}
\includegraphics[width=12.cm]{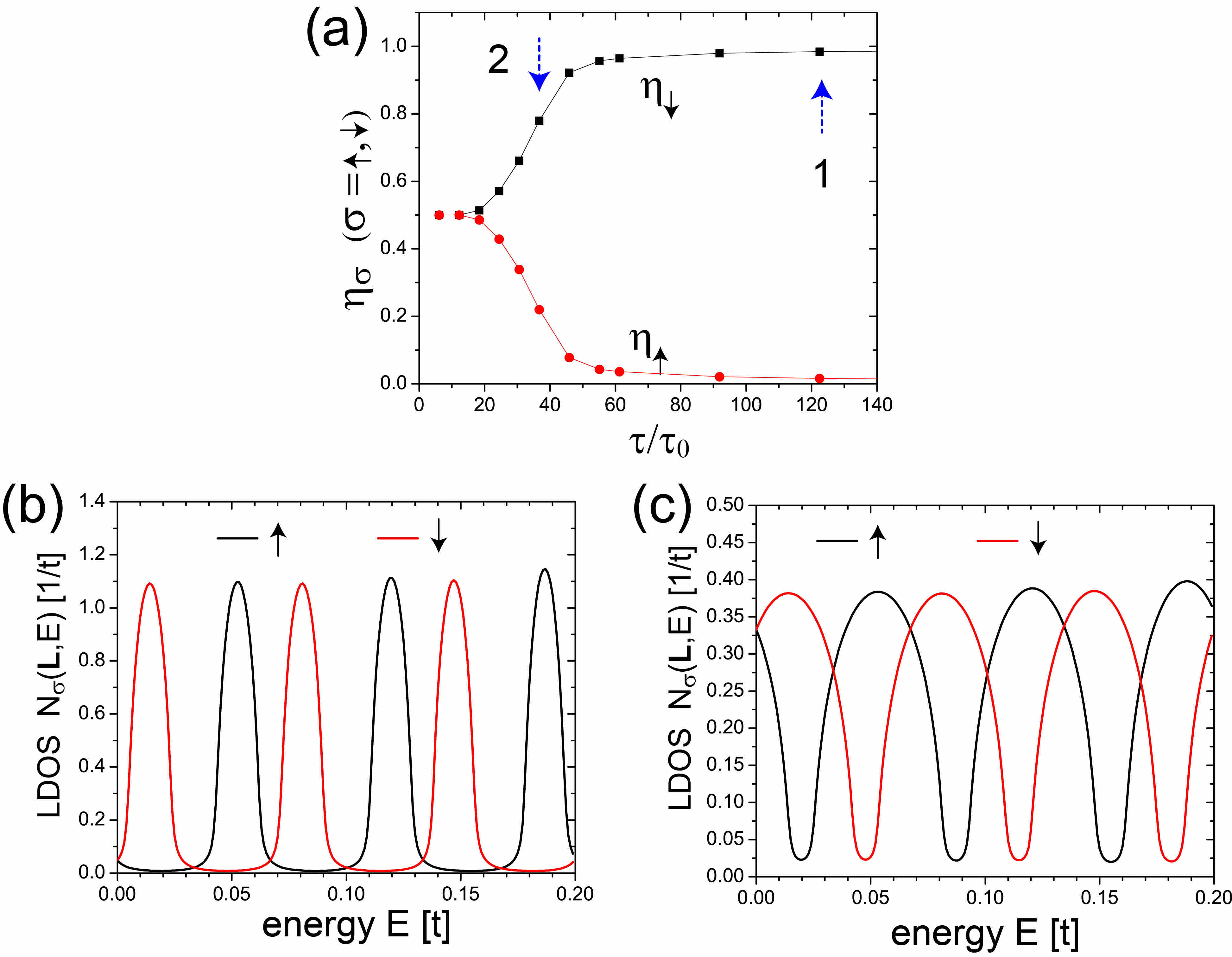}
\caption{TI with a magnetic defects of Ising-symmetry with $J_{z} S = 5t$, $J_\pm S=0$, and $t_{l}=0.1t$.  (a) $\eta_\sigma$ as a function of $\tau/\tau_0$ (see text) obtained for $V_g=0.014t/e$. (b) and (c) $N_\sigma({\bf L},E)$ for values of $\tau/\tau_0$ indicated by arrows 1 and 2 in (a), respectively.  } \label{fig:Dephasing_Jz}
\end{center}
\end{figure}

We begin by investigating the effect of dephasing on the currents' spin polarization in a TI with a magnetic defect of Ising symmetry (i.e., $J_z S \not = 0, J_\pm S= 0$).  In Fig.~\ref{fig:Dephasing_Jz}(a) we present the current's spin polarization $\eta_\sigma$ as a function of $\tau/\tau_0$, where $\tau$ is the electronic lifetime extracted from the energy width of the peaks in the LDOS, and $\tau_0 = \hbar /t $ is the hopping time between nearest neighbor sites. In general, $\tau \sim 1/\sqrt{\gamma}$ with $\gamma = 2 g^2 k_B T/\omega_0$ reflecting the strength of the electron-phonon coupling [see Eq.(\ref{eq:sigma})]. The current's spin polarization remains unaffected by dephasing as long as the corresponding energy width of the spin-split edge states, $\Gamma = \hbar / \tau$, is smaller than their energy splitting $\Delta E$ (see Fig.\ref{fig:Dephasing_Jz}(b)), corresponding to a value of $\tau/\tau_0$ denoted by  arrow 1 in Fig.\ref{fig:Dephasing_Jz}(a)). Once the lifetime of the spin-split edge states becomes sufficiently short such that they start to overlap in energy (see Fig.\ref{fig:Dephasing_Jz}(c), corresponding to a value of $\tau/\tau_0$ denoted by  arrow 2 in Fig.\ref{fig:Dephasing_Jz}(a)), the spin polarization begins to decrease. This is expected since dephasing reverses the effect of the magnetic defect: the latter separates the \su{} and \sd{} states in energy, while the former increases the energy width of these states and thus leads to an overlap in energy.

We next investigate the effects of dephasing in a TI containing two magnetic defects of $xy$-symmetry (i.e., $J_\pm S \not = 0, J_z S= 0$).
\begin{figure}[h]
 \begin{center}
\includegraphics[width=11cm]{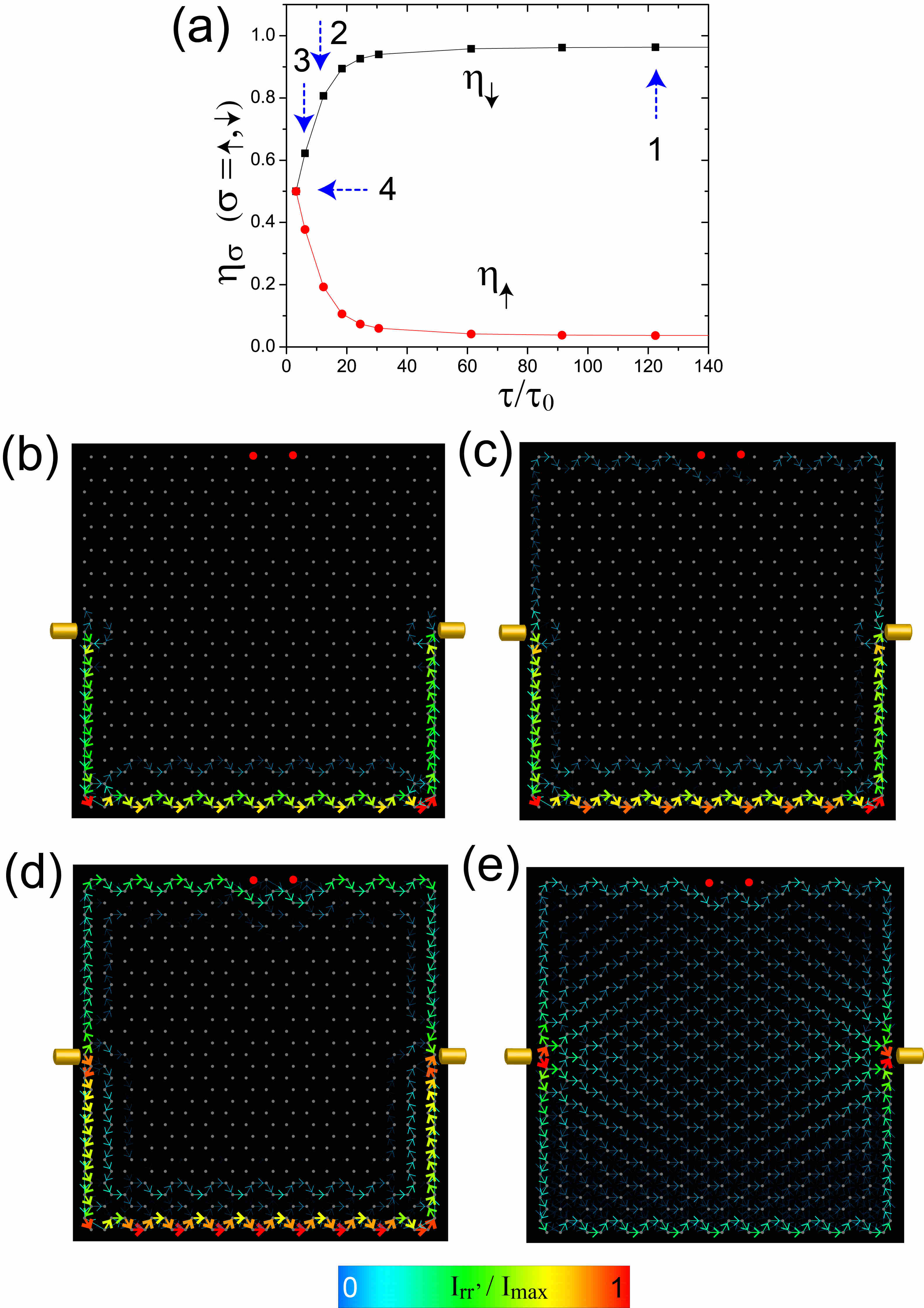}
\caption{TI containing two magnetic defects of $xy$-symmetry (see filled red circles in (b)) with $J_{\pm} S = 5t$, $J_z S= 0$, and $t_{l}=0.5t$. (a) $\eta_\sigma$ as a function of $\tau/\tau_0$ (see text) for $V_g=0.0175t/e$. (b) - (e) $I^c_{\vect{r} \vect{r}'}$ for the same $V_g$ and the values of $\tau/\tau_0$ indicated by arrows 1 - 4 in (a) respectively.} \label{fig:Dephasing_Jxy}
\end{center}
\end{figure}
In Fig.~\ref{fig:Dephasing_Jxy}(a) we present the spin polarization $\eta_\sigma$ as a function of $\tau/\tau_0$. In such a system, the spin polarization begins to decrease when $\gamma$ becomes sufficiently large such that the edge states start to overlap in energy, and vanishes when the edge states begin to hybridize with the bulk states, which implies that the topological nature of the system is destroyed. This is evidenced by the spatial patterns of the charge current shown in Figs.~\ref{fig:Dephasing_Jxy}(b) - (e). At large values of the lifetime $\tau$ (see arrow 1 in Fig.~\ref{fig:Dephasing_Jxy}(a)), the spatial pattern is similar to that obtained for $\gamma = 0$, see Fig.~2(d) of Ref. \cite{VanDyke2016}. Concomitant with a decrease in the spin polarization, one finds that the current carried by the edge states extends further into the bulk of the TI (see Figs.~\ref{fig:Dephasing_Jxy}(c) and (d), corresponding to arrows 2 and 3 in Fig.~\ref{fig:Dephasing_Jxy}(a), respectively), circumventing the effect of the magnetic defect.  This indicates the onset of overlap in energy  between the edge states, since higher energy edge states possess a larger decay length along the zig-zag edge, implying a penetration of the edge currents into the bulk of the system, as discussed in Sec.~\ref{sec:evol}. Upon further decreasing $\tau$, the spin polarization eventually vanishes (i.e., $\eta_\uparrow = \eta_\downarrow = 0.5$), while at the same time, a substantial portion of the current flows through the center of the TI, as shown in Fig.~\ref{fig:Dephasing_Jxy}(e), corresponding to arrow 4 in Fig.~\ref{fig:Dephasing_Jxy}(a). This indicates a significant hybridization between bulk and edge states via the electron-phonon coupling, and consequently, a destruction of the topological nature of the system arising from dephasing \cite{Qi2011}. Note, however, that the spin polarization is significantly more robust against dephasing in the case of magnetic defects with $xy$-symmetry, than in the present of magnetic defects with Ising-symmetry.

\section{Symmetries of the spin current}
\label{sec:symmetries}

Using the spin-polarized currents in  finite two-dimensional TIs for applications in quantum computation or spin electronics will require us to identify the symmetry properties of a current's spin polarization, or equivalently, of the spin current, $i_s = \eta_\uparrow - \eta_\downarrow$, under various transformations such as a reversal of the bias or gate voltages, spatial reflections or rotations around points or lines, and sign changes of the magnetic defect's scattering potential or of the spin-orbit coupling. We will discuss these symmetry properties, which differ for magnetic defects with Ising- and $xy$-symmetry, in the following.

\subsection{Symmetries of the spin current in the presence of magnetic defects with Ising symmetry}

\begin{figure}[h]
 \begin{center}
\includegraphics[width=9cm]{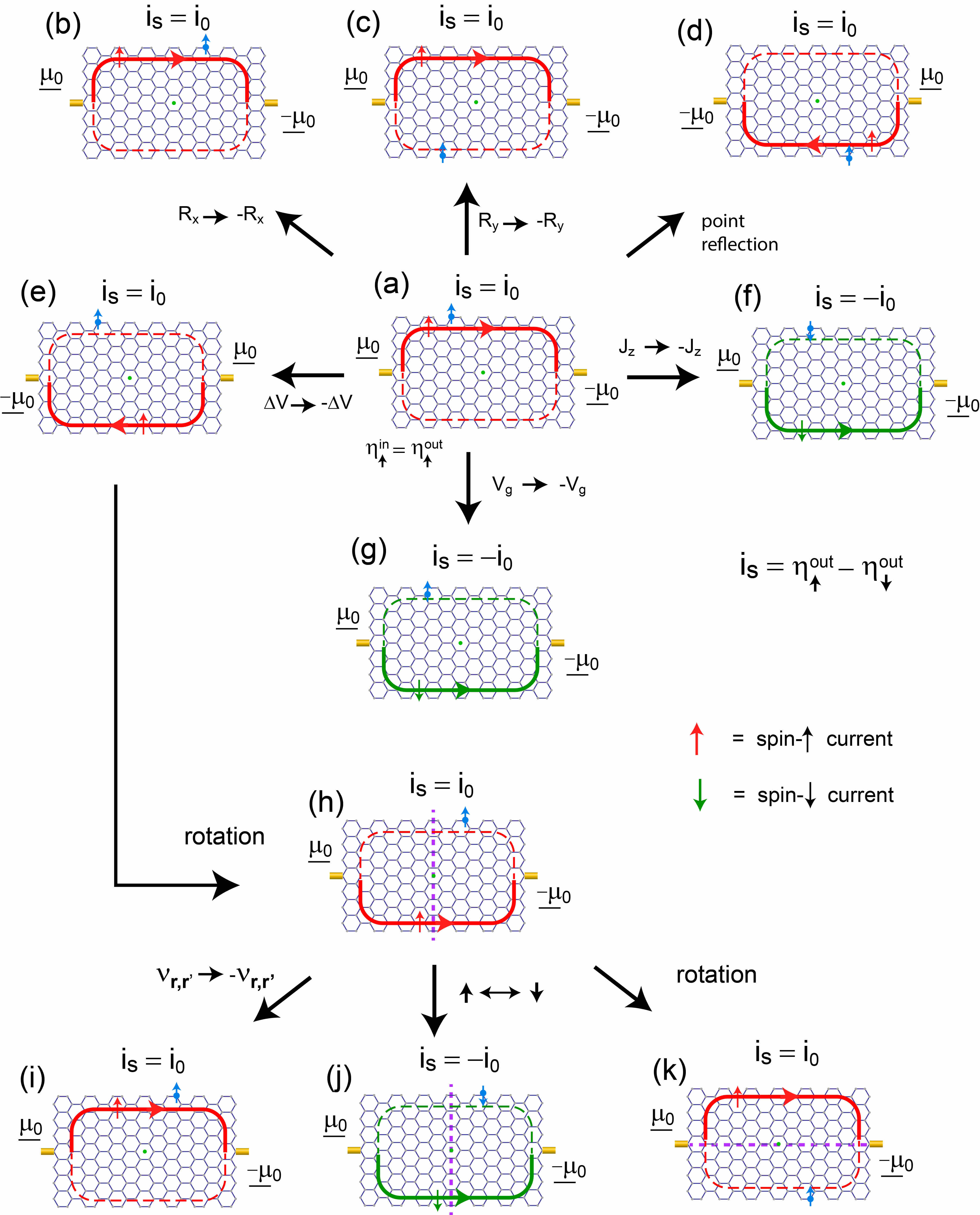}
\caption{TI with a magnetic defect of Ising-symmetry $(J_z S \not = 0)$ located at ${\bf R}=(R_x,R_y)$. The spatial patterns of $I^\uparrow$ and  $I^\downarrow$ are schematically represented by red and green lines, respectively, and $\mu_{L,R} = \pm \mu_0$.
(a) $I^\uparrow$ obtained for a given set of $J_z$, $\Delta V$, $V_g$ with spin current $i_s = i_0>0$.
(b) $I^\uparrow$ obtained from (a) under $R_x \rightarrow -R_x$, with $i_s = i_0$.
(c) $I^\uparrow$ obtained from (a) under $R_y \rightarrow -R_y$, with $i_s = i_0$.
(d) $I^\uparrow$ obtained from (a) under reflection at the TI's center point, with $i_s = i_0$.
(e) $I^\uparrow$ obtained from (a) under $\Delta V \rightarrow - \Delta V$, with $i_s = i_0$.
(f) $I^\downarrow$ obtained from (a) under $J_z \rightarrow -J_z$, with $i_s = -i_0$.
(g) $I^\downarrow$ obtained from (a) under $V_g \rightarrow -V_g$, with $i_s = -i_0$.
(h) $I^\uparrow$ obtained from (e) under rotation around a vertical axis through the TI's center, with $i_s=i_0$.
(i) $I^\uparrow$ obtained from (h) under $\nu_{\bf r,r^\prime} \rightarrow -\nu_{\bf r,r^\prime}$, with $i_s = i_0$.
(j) $I^\downarrow$ obtained from (h) under $\uparrow \leftrightarrow \downarrow$, with $i_s = - i_0$.
(k) $I^\downarrow$ obtained from (h) under rotation around a horizontal axis through the TI's center, with $i_s=i_0$.} \label{fig:Jzsymm}
\end{center}
\end{figure}
For defects with Ising-symmetry $(J_z S\not = 0, J_\pm S= 0)$, the transformation properties of the spin current under various symmetry operations are shown in Fig.~\ref{fig:Jzsymm}. As discussed above, such defects lift the degeneracy of the Kramers doublet of edge states, allowing one to select a spin-polarized state for charge transport via gating. In Fig.~\ref{fig:Jzsymm}(a), we consider charge transport through a \su\ state, and schematically depict the spatial \su\ current patterns for a given set of $J_z$, $\Delta V$, $V_g$, giving rise to a spin current $i_s =\eta_\uparrow - \eta_\downarrow = i_0>0$. This current pattern is similar to that shown in Figs.~1(c) and 2(b) of Ref. \cite{VanDyke2016}, with the backflow branch indicated by a narrower dashed line. The magnetic defect (denoted by a blue arrow) is located at ${\bf R}=(R_x,R_y)$, with the origin being at the center of the TI.  The spin current is invariant under $R_x \rightarrow -R_x$ (see Fig.~\ref{fig:Jzsymm}(b)), $R_y \rightarrow -R_y$ (see Fig.~\ref{fig:Jzsymm}(c)), and point reflection (of the entire system, including the leads) at the center point of the TI (see Fig.~\ref{fig:Jzsymm}(d)), since the electronic structure of the TI, and in particular, the splitting of the Kramers doublet of edge states, remains unchanged. Similarly, the spin current remains invariant under bias reversal $\Delta V \rightarrow -\Delta V$ (see Fig.~\ref{fig:Jzsymm}(e)) since the same \su\ state is still utilized for charge transport, albeit the direction of charge flow has been reversed. On the other hand, a sign change of the scattering potential $J_z \rightarrow -J_z$ (see Fig.~\ref{fig:Jzsymm}(f)) exchanges the energy shifts of the \su\ and \sd\ states, such that for fixed $V_g$, one now utilizes a \sd\ polarized state for charge transport, thus leading to a change in the sign of the spin-current. Similarly, since the energy shifts for the \su\ and \sd\ bands arising from Ising-type magnetic defects are just opposite in sign, a reversal of the gate voltage $V_g \rightarrow -V_g$ (see Fig.~\ref{fig:Jzsymm}(g)) implies that one now accesses a \sd\ polarized state for charge transport, leading to a reversal in the sign of the spin current $i_s$.

To demonstrate how all of these different transformations and the resulting symmetries of the spin current are related, we start from the case shown in Fig.\ref{fig:Jzsymm}(e). When the system is rotated around the center axis (indicated by a vertical dashed line in Fig.\ref{fig:Jzsymm}(h)), one arrives at a TI (Fig.\ref{fig:Jzsymm}(h)) in which the spin current remains unchanged. However, this rotation changes the overall sign of $\nu_{\bf r,r^\prime}$, see Eq.(\ref{eq:fullH}), such that this TI's spin-orbit coupling has a sign opposite to that of the original system (Fig.\ref{fig:Jzsymm}(e)). To restore the sign of the original model, one can proceed in three different ways, as shown in Figs.\ref{fig:Jzsymm}(i) - (k). First, one can change the sign of $\nu_{\bf r,r^\prime}$, i.e., $\nu_{\bf r,r^\prime} \rightarrow -\nu_{\bf r,r^\prime}$ which changes the direction of current flows, as shown in Fig.\ref{fig:Jzsymm}(i), but leaves the spin current invariant. This TI is identical to the one shown in Fig.\ref{fig:Jzsymm}(b). Second, one can exchange the spin degrees of freedom, i.e., $\uparrow \leftrightarrow \downarrow$, for the conduction electrons in the TI as well as for the magnetic defect, which changes the sign of the spin current. The resulting TI shown in Fig.\ref{fig:Jzsymm}(j) is related to the one shown in Fig.\ref{fig:Jzsymm}(f) via $R_x \rightarrow -R_x$, which leaves the spin-current unchanged. Finally, one can rotate the system around the horizontal axis (dashed line in Fig.\ref{fig:Jzsymm}(k)), which leaves the spin-current unchanged, but moves the defect to the lower side of the TI as shown in Fig.\ref{fig:Jzsymm}(k). This TI is related to those shown in Figs.\ref{fig:Jzsymm}(b) and (c)  via $R_y \rightarrow -R_y$ and $R_x \rightarrow -R_x$, respectively.

\subsection{Symmetries of the Spin Current in the presence of magnetic defects with $xy$-symmetry}

\begin{figure}[h]
 \begin{center}
\includegraphics[width=8.5cm]{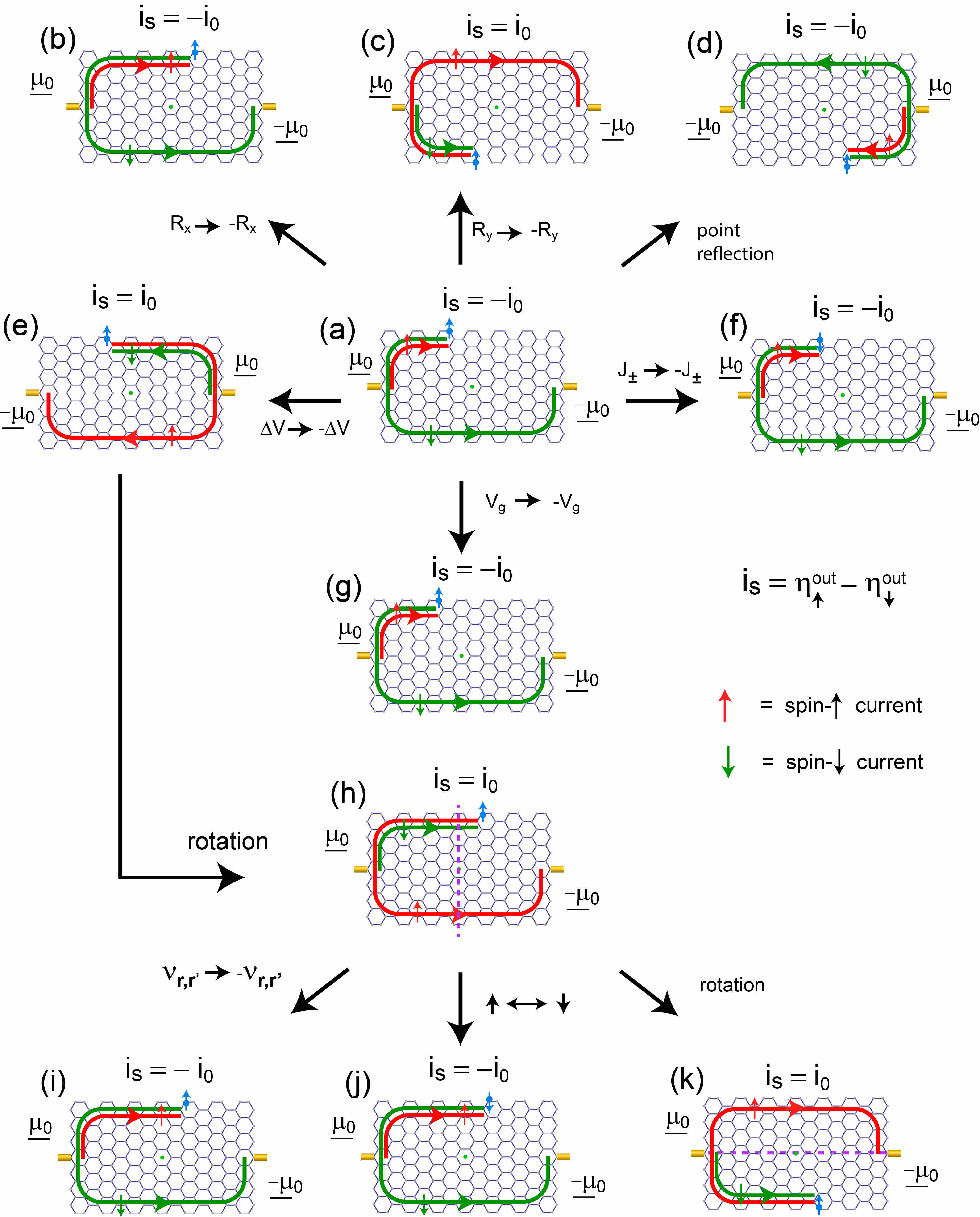}
\caption{TI with a magnetic defect of $xy$-symmetry $(J_\pm S \not = 0)$ at ${\bf R}=(R_x,R_y)$, and $\mu_{L,R} = \pm \mu_0$. The \su and \sd current patterns are schematically presented by  red and green lines, respectively.
(a) $I^\sigma$ for a given set of $J_\pm$, $\Delta V$, $V_g$, yielding a spin current $i_s = -i_0<0$.
(b) $I^\sigma$  obtained from (a) under $R_x \rightarrow -R_x$ with $i_s = -i_0$.
(c) $I^\sigma$ obtained from (a) under $R_y \rightarrow -R_y$ with $i_s = i_0$.
(d) $I^\sigma$ obtained from (a) under reflection at the center point of the TI, with $i_s = -i_0$.
(e) $I^\sigma$ obtained from (a) under $\Delta V \rightarrow - \Delta V$, with $i_s = i_0$.
(f) $I^\sigma$ obtained from (a) under $J_\pm \rightarrow - J_\pm$, with $i_s = -i_0$.
(g) $I^\sigma$ obtained from (a) under $V_g \rightarrow -V_g$, with $i_s = -i_0$.
(h) $I^\sigma$ obtained from (e) under rotation around a vertical axis through the center of the TI, with $i_s=i_0$.
(i) $I^\sigma$ obtained from (h) under $\nu_{\bf r,r^\prime} \rightarrow -\nu_{\bf r,r^\prime}$, with $i_s = -i_0$.
(j) $I^\sigma$ obtained from (h) under $\uparrow \leftrightarrow \downarrow$, with $i_s = -i_0$.
(k) $I^\sigma$ obtained from (h) under rotation around a horizontal axis through the center of the TI, with $i_s=i_0$.
} \label{fig:Jxysymm}
\end{center}
\end{figure}

The transformation properties of the spin current for a TI in the presence of a magnetic defect with $xy$-symmetry are shown in Fig.~\ref{fig:Jxysymm}. In Fig.~\ref{fig:Jxysymm}(a), we schematically depict the spatial \su\ (red) and \sd\ (green) current patterns for a given set of $J_\pm$, $\Delta V$, $V_g$, giving rise to the spin current $i_s = - i_0<0$. These current patterns are similar to those shown in Fig.~\ref{fig:1magdef}(b) and (c) (or in Fig.~2(e) and (f) of Ref. \cite{VanDyke2016}), where the backflow branch is absent due to a larger coupling $t_{l}$ to the leads. The magnetic defect is located at ${\bf R}=(R_x,R_y)$ (as indicated by a blue arrow), with the origin being at the center of the TI. Under the transformation $R_x \rightarrow -R_x$ (Fig.~\ref{fig:Jxysymm}(b)) the spin current remains invariant, since the defect is still located in the path of the \su\ current and the electronic structure of the TI remains unchanged. However, under the transformation $R_y \rightarrow -R_y$ (Fig.~\ref{fig:Jxysymm}(c)) the sign of the spin current changes since the defect is now located in the path of the \sd\ current, and hence scatters predominantly \sd\ electrons into the  \su\ band, with a concomitant change in the spatial \su\ and \sd\ current patterns schematically shown in Fig.~\ref{fig:Jxysymm}(c). On the other hand, under point reflection (of the entire system, including the leads) at the center point of the TI (Fig.~\ref{fig:Jxysymm}(d)), the spin current remains unchanged since the defect is still located in the path of the \su\ current. In contrast, under bias reversal $\Delta V \rightarrow -\Delta V$ (Fig.~\ref{fig:Jxysymm}(e)), the defect is located in the path of the \sd\ current, leading to a sign change of $i_s$. Moreover, a sign change of the scattering potential, $J_{\pm} \rightarrow -J_{\pm}$ (Fig.~\ref{fig:Jxysymm}(f)) leaves the spin current unchanged since corrections to the \su\ and \sd\ electronic structure arising from the defect scattering contain only even powers of the scattering potential. Similarly, the spin current remains unchanged under reversal of the gate voltage $V_g \rightarrow -V_g$ (Fig.~\ref{fig:Jxysymm}(g)) since the scattering potential does not break the particle-hole symmetry of the \su\ and \sd\ bands.

To exemplify the relation between all of these different transformations and the resulting symmetries of the spin current, we start from the case shown in Fig.\ref{fig:Jxysymm}(e). When the system is rotated around the center axis (indicated by a vertical dashed line in Fig.\ref{fig:Jxysymm}(h)), one arrives at a TI (Fig.\ref{fig:Jxysymm}(h)) in which the spin current is unchanged. However, this rotation changes the overall sign of $\nu_{\bf r,r^\prime}$, see Eq.(\ref{eq:fullH}), such that this TI's spin-orbit coupling has a sign  opposite to that of the original system (Fig.\ref{fig:Jxysymm}(e)). To restore the sign of the original model, one can proceed in three different ways, shown in Figs.\ref{fig:Jxysymm}(i) - (k). First, one can change the sign of $\nu_{\bf r,r^\prime}$, i.e., $\nu_{\bf r,r^\prime} \rightarrow -\nu_{\bf r,r^\prime}$ (Fig.\ref{fig:Jxysymm}(i)) which reverses the direction of charge flow. As a result, the sign of the spin current is changed since the defect is now located in the path of the \su\ current. This TI is identical to the one shown in Fig.\ref{fig:Jxysymm}(b). Second, exchanging the spin degrees of freedom, i.e., $\uparrow \leftrightarrow \downarrow$, for the conduction electrons in the TI as well as for the magnetic defect leads to a sign change of the spin current, since the defect is now located in the path of the \su\ current, as shown in Fig.\ref{fig:Jxysymm}(j). This TI is related to the one shown in Fig.\ref{fig:Jxysymm}(b) via $J_\pm \rightarrow -J_\pm$ and to the one shown in Fig.\ref{fig:Jxysymm}(f) via $R_x \rightarrow -R_x$. Finally, one can rotate the system around the horizontal axis (dashed line in Fig.\ref{fig:Jxysymm}(k)), which leaves the spin current unchanged, but moves the defect to the lower side of the TI as shown in Fig.\ref{fig:Jxysymm}(k). This TI is related to the one shown in Fig.\ref{fig:Jxysymm}(c) via $R_x \rightarrow -R_x$.

\section{Creation of Localized Magnetic Fields: interplay of topology and symmetry of the magnetic defects}
\label{sec:magfield}

\begin{figure}[t]
 \begin{center}
\includegraphics[width=5cm]{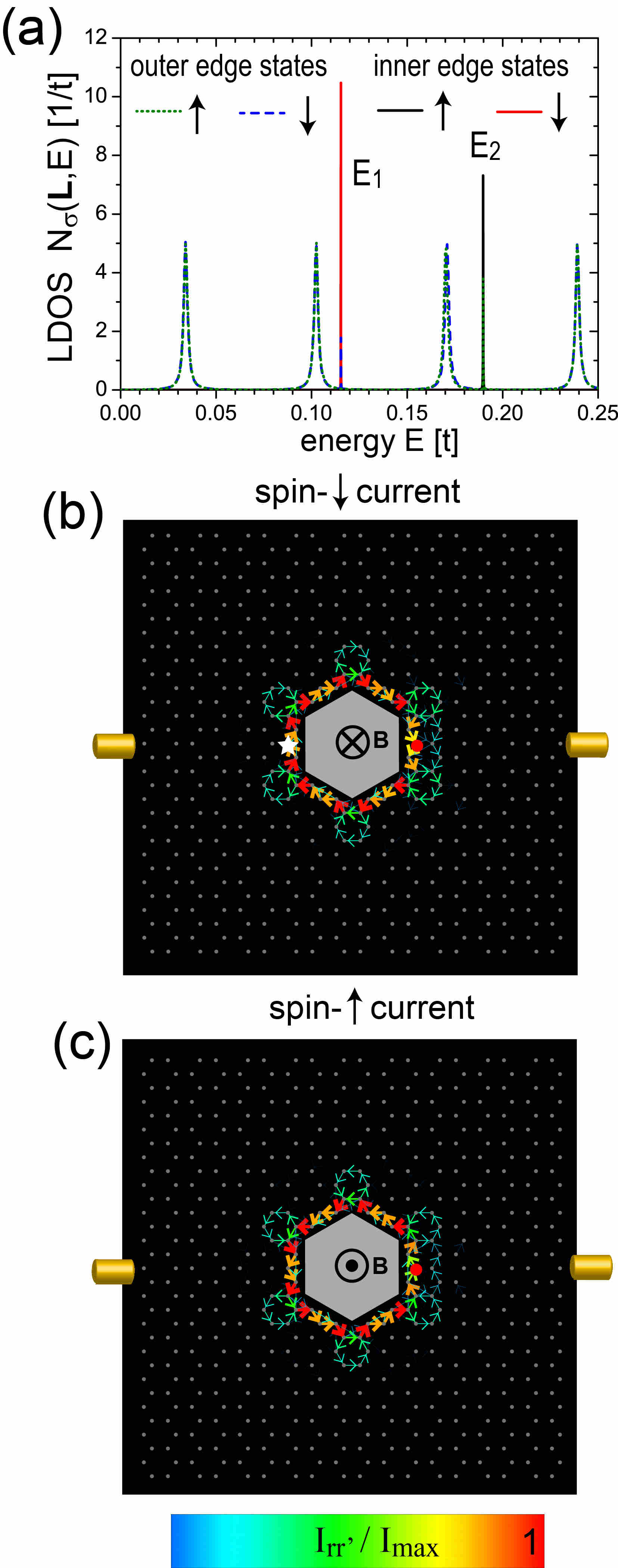}
\caption{TI with a hole in its center (rendered as a shaded gray area in (b), (c)) with $t_l=0.1t$ and a magnetic defect of Ising symmetry ($J_z S=5t$) (see filled red circle in (c)). (a) $N_\sigma(E)$ at ${\bf L}$ (for the outer edge states) and at the site denoted by a white star (see (b)) for the inner edge states.  (b) $I_{\vect{r} \vect{r}'}^\downarrow$ for $V_g=E_1/e=0.1153t/e$ (see (a)). The induced magnetic field, $B$, points into the plane. (c) $I_{\vect{r} \vect{r}'}^\uparrow$ for $V_g=E_2/e=0.1898t/e$. The induced magnetic field, $B$, points out of the plane.} \label{fig:TI_holes}
\end{center}
\end{figure}

The insight into the question of how the interplay between the topology and symmetry of the magnetic defects allows us to create spin-polarized current, also enables us to propose intriguing quantum phenomena that are based on this insight. An example is the ability to design spatially highly localized magnetic fields using holes created in the interior of a TI, as shown in Fig. \ref{fig:TI_holes}. A magnetic defect of Ising symmetry placed near such a hole leads to a much larger splitting of the (inner) \su\ and \sd\ edge states surrounding the hole than of the outer perimeter edge states (Fig. \ref{fig:TI_holes}), due to a much weaker coupling of the magnetic defect to the latter. Conversely, the inner edge states are much more weakly coupled to the leads, resulting in their much narrower width in energy. By gating the TI to the energy of one of the spin-resolved inner edge states, one creates either clockwise (Fig. \ref{fig:TI_holes}) or counter-clockwise (Fig. \ref{fig:TI_holes}) circulating currents, leading to magnetic fields of opposite direction in the center of the hole. Such spatially highly localized magnetic fields, whose magnitude is controlled by the hole size, are of great interest for the realization of nanoscale nuclear magnetic resonance \cite{Mamin2013}. Note that the weak hybridization between the inner and outer edge states implies that these interior magnetic fields can be created with only a negligible net charge current flowing through the TI (Fig. \ref{fig:TI_holes}).

\section{Conclusions}
\label{sec:concl}

In this article we have investigated the effects of defects of various natures and dephasing arising from an electron-phonon interaction on the electronic structure and transport properties of two-dimensional topological insulators. We have demonstrated how the spatial current patterns evolve between the topologically protected edge states and the bulk states. Moreover, we have shown that elastically and inelastically scattering defects that do not break the time-reversal symmetry of the TI have qualitatively different effects on the TI's electronic structure and transport properties. In particular, we found that an inelastically scattering defect (for example, a phonon mode inside a molecule) suppresses the conductance of the TI even when it preserves the TI's time reversal symmetry, as it leads to a hybridization of edge and bulk states. We also showed that the recently predicted ability to create highly spin-polarized currents in a TI by breaking its time-reversal symmetry via magnetic defects or in magnetic heterostructures, is robust under changes over wide range of materials parameters including the magnetic scattering strength, size and aspect ratio of the TI, width of the leads, or the strength $\Lambda_{SO}$ of spin-orbit coupling strength. This effect is also robust against dephasing induced by the interaction with phonons or the inclusion of a Rashba spin-orbit interaction, as long as these two interactions do not destroy the global topological nature of the TI. We explored the symmetry properties of the spin polarization under various spatial transformations of the system, as well as reversal of the bias and gate voltages, which is of great importance for the manipulation of spin-polarized currents in future devices. We proposed that the interplay between topology and the symmetry of the magnetic defects can be employed to create novel quantum phenomena, such as the creation of highly localized magnetic fields around interior holes in a TI, whose direction can be tuned through the gate voltage.  This prediction adds another example to the growing list of applications for topological insulators.

We note in passing that effects of potential and magnetic defects on the spatial form of current patterns in 2D TIs, similar to the ones reported in Ref.\cite{VanDyke2016} and above, were subsequently also discussed in Refs. \cite{Dang2015,Dang2016b}.

\section{Acknowledgement}
We would like to thank P. Lee, J. Van Den Brink and M. Vojta for helpful discussions.  This work was supported by the U. S. Department of Energy, Office of Science, Basic Energy Sciences, under Award No. DE-FG02-05ER46225.

\end{document}